\documentclass[preprint,12pt]{elsarticle}

\usepackage{graphicx,dcolumn,bm,xcolor,microtype,hyperref,multirow,amscd,amsmath,amssymb,amsfonts,physics}
\usepackage[version=4]{mhchem}
\usepackage{algorithmicx,algorithm,algpseudocode}
\usepackage{ulem}

\newcolumntype{H}{>{\setbox0=\hbox\bgroup}c<{\egroup}@{}}

\algnewcommand\algorithmicswitch{\textbf{switch}}
\algnewcommand\algorithmiccase{\textbf{case}}
\algnewcommand\algorithmicassert{\texttt{assert}}
\algnewcommand\Assert[1]{\State \algorithmicassert(#1)}
\algdef{SE}[SWITCH]{Switch}{EndSwitch}[1]{\algorithmicswitch\ #1\
\algorithmicdo}   {\algorithmicend\ \algorithmicswitch}
\algdef{SE}[CASE]{Case}{EndCase}[1]{\algorithmiccase\ 
#1}{\algorithmicend\  \algorithmiccase} \algtext*{EndSwitch}
\algtext*{EndCase}
\algrenewcommand{\algorithmiccomment}[1]{$\triangleright$
\textit{#1}}
\algnewcommand{\algorithmicgoto}{\textbf{go to}}%
\algnewcommand{\Goto}[1]{\algorithmicgoto~\ref{#1}}%

\biboptions{numbers,sort&compress}

\hypersetup{pdfauthor={Name}}

\journal{Advances in Quantum Chemistry}

\begin{document}

\begin{frontmatter}

\title{Advances in Approximate Natural Orbital Functionals: From Historical Perspectives to Contemporary Developments}
\author{Mario Piris$^{1,2}$}

\ead{mario.piris@ehu.eus}

\address{$^{1}$Kimika Fakultatea, Euskal Herriko Unibertsitatea (UPV/EHU) 
and Donostia International Physics Center (DIPC), P.K. 1072, 20080 Donostia, Euskadi, Spain.}
\address{$^{2}$Basque Foundation for Science (IKERBASQUE), 48013 Bilbao, Euskadi, Spain.}

\begin{abstract}
This chapter provides a comprehensive review of fundamental concepts related to approximate natural orbital functionals (NOFs), emphasizing their significance in quantum chemistry and physics. Focusing on fermions, the discussion excludes considerations of finite temperature and systems with a variable number of particles. The theoretical foundation for approximate NOFs is laid out, with a particular emphasis on functional N-representability. Various two-index reconstructions for the two-particle reduced density matrix (2RDM) are introduced, accompanied by discussions on challenges. The analysis delves deeply into NOFs grounded in electron pairing, specifically focusing on PNOF5, PNOF7, and the Global NOF, a more versatile approach addressing both static and dynamic electron correlation components. The extension of NOFs to multiplets while conserving total spin is presented, and the availability of open-source implementations like \href{http://github.com/DoNOF}{DoNOF} and its associated programs is highlighted. A detailed overview of optimization procedures for single-point calculations is provided. Sections on geometry optimization and ab initio molecular dynamics, closely connected to the availability of analytical gradients in NOF theory, are presented. The chapter concludes with the extension of NOFs to both charged and excited states.
\end{abstract}


\end{frontmatter}

\section{\label{sec:Intro}Introduction}

Natural orbitals (NOs), introduced by Per-Olov Löwdin in 1955 \cite{Lowdin1955a}, represent a pivotal concept in quantum chemistry and quantum physics. These orbitals provide a comprehensive framework for understanding the electronic structure of atoms, molecules and solids; allowing for accurate descriptions of their electronic distributions and interactions. Indeed, most \textit{ab initio} methods of electronic structure are based on the expansion of approximate wavefunctions into a series of antisymmetrized products of one-particle functions that we call orbitals. Within these methods, an efficient selection of the orbitals is fundamental, and it is at this moment that NOs arise that lead to the diagonalization of the one-particle reduced density matrix (1RDM).

Löwdin and Shull showed \cite{Lowdin1955d} that NOs could be used to express the two-electron wavefunction in the simplest way, that is, with the fewest number of configurations. The successful application of this representation to describe the ground states of both the helium atom \cite{Shull1959} and the hydrogen molecule \cite{Shull1959a} was the source of inspiration that sparked widespread interest in NOs and led to important results in systems with more than two electrons \cite{McWeeny1960, Smith1968}. The general properties and uses of NOs have been a subject of extensive investigation in the past \cite{Davidson1972} and remain a subject of significant interest in contemporary research \cite{Cioslowski2021, Cioslowski2023}, along with local descriptors of dynamic and nondynamic correlation that rely on them \cite{Ramos-Cordoba2017, Xu2023}.

The importance of NOs increased with the emergence of 1RDM functional theory (1RDMFT) \cite{Gilbert1975, Levy1979, Valone1980}, since the vast majority of 1RDM functionals introduced to date are formulated in the NO representation. Such functionals are, by definition, natural orbital functionals (NOFs). For a complete historical overview of the formulation and evolution of NOF theory (NOFT) up to 2006, readers are referred to the author's previous review article \cite{Piris2007}. 

In this review, the aim is to provide insight into some recent advances in NOFT that have occurred since then, recognizing that 1RDMFT continues to be a vibrant field of research \cite{Schilling2019, Cioslowski2019, Rodriguez-Mayorga2022, Senjean2022, Liebert2023}. The focus is on fermions, specifically electrons subject to Coulomb interactions. Of note, significant progress has been made in establishing the fundamental principles of 1RDMFT for bosons \cite{Benavides-Riveros2020,Liebert2021}. At present, the primary challenge revolves around the establishment of reliable NOF approximations for bosons. Nevertheless, it is heartening to observe that significant progress has already been achieved in tackling this task \cite{Schmidt2021}. 

We consider only systems with an integer number N of electrons that are at zero temperature. 1RDMFT at finite temperatures can be found elsewhere \cite{Baldsiefen2015, Baldsiefen2017, Giesbertz2018, Sutter2023}. In the realm of 1RDMFT, systems featuring a variable number of particles have also been an area of interest \cite{Alcoba2007, Lathiotakis2010, Head-Marsden2015, Piris2015, Acke2023}. A notably comprehensive study by Giesbertz and Ruggenthaler \cite{Giesbertz2019} has addressed 1RDMFT at elevated temperatures and with variable particle numbers, specifically in the context of grand canonical ensembles encompassing both fermions and bosons.

This chapter is also restricted to the stationary NOFT. The time-dependent evolution of the 1RDM has been investigated using simplified methodologies such as the adiabatic approximation and linear response techniques, showing favorable results, especially in simple systems like the hydrogen molecule \cite{Pernal2007a, Pernal2007, Giesbertz2008, Giesbertz2010a, Apple2010, Brics2013, Benavides-Riveros2019}. The establishment of a robust framework for dynamic NOFT remains an ongoing and open challenge \cite{Pernal2016}. Nonetheless, it is important to note that in this review we will delve into recently developed formalisms designed to address excited electronic states in the context of the equation-of-motion method \cite{Rowe1968}, as well as the molecular dynamics of nuclei in the Born-Oppenheimer (BO) approximation. The latter approach enables us to monitor the temporal evolution of the 1RDM by solving the stationary problem for every instantaneous configuration of nuclei.

The existence of the 1RDM functional itself does not provide the means for its construction, although for some model systems the Levy construction \cite{Levy1979} has allowed to discover the explicit form of the exact functional in terms of the 1RDM \cite{Tows2011, Cohen2016, Muller2018}. The main problem with the constrained search is generating all suitable many-electron wavefunctions (pure states) or N-electron density matrices (mixed states). Hence, the constrained search formulation is not suitable for computational purposes. This limitation has prompted the development of approximate functionals designed for practical applications.

In the context of the systems of interest, the Hamiltonian operator is composed of both one-electron and two-electron operators. Consequently, the electronic system's energy can be exactly determined by having access to the one- and two-particle reduced density matrices, denoted as 1RDM and 2RDM, and represented as $\Gamma$ and D, respectively. The non-interacting part of the Hamiltonian corresponds to the one-particle operator and already exhibits an explicit dependence on the 1RDM. This underscores a fundamental advantage of the 1RDM formulation, as the kinetic energy is explicitly defined and does not require the construction of a functional, unlike the case with a density functional. Therefore, our only objective is to reconstruct the electron-electron potential energy, denoted as V$_{ee}$, which constitutes the segment of the energy explicitly dependent on the 2RDM, and express it in terms of the 1RDM.

A typical approach involves employing the exact functional expression of V$_{ee}$[D], using a 2RDM constructed via a reconstruction functional, D[$\Gamma$], to formulate a V$_{ee}$ functional. Evidently, this electron-electron potential energy is not the exact functional of the 1RDM, the only one explicitly dependent on $\Gamma$. Consequently, an approximate functional maintains its dependence on the 2RDM \cite{Donnelly1979}. A significant drawback of this dependence is the emergence of the functional N-representability problem \cite {Ludena2013, Piris2018d}. This issue necessitates that the 2RDM, reconstructed in terms of the 1RDM, must adhere to the same N-representability conditions as those applied to the unreconstructed 2RDMs \cite{Mazziotti2012}. Failing to meet these conditions could result in the absence of a compatible N-electron fermionic system for the energy functional. In essence, we are no longer purely engaged in the 1RDM functional theory but rather in an approximate one-particle theory, where the 2RDM continues to play a significant, albeit hidden, role. The N-representability of the approximate functional depends on the N-representability of the reconstructed 2RDM.

Unfortunately, a prevailing assumption has been that approximate functionals do not encounter N-representability issues, primarily because the N-representability conditions for the 1RDM \cite{Coleman1963} were perceived as self-sufficient. Some energy expressions, like the one proposed by M\"{u}ller \cite{Muller1984}, appear to be appropriately formulated in terms of the 1RDM and may even yield reasonably accurate results for specific systems. However, these functionals transgress N-representability conditions as fundamental as the requirement for the 2RDM to be positive semidefinite. This leads to the realization that many of the approximate functionals currently employed \cite{Sharma2008, Marques2008, Rohr2008} are not N-representable \cite{RodriguezMayorga2017, Mitxelena2017a, Mitxelena2018}.

The discovery \cite{Klyachko2006, Altunbulak2008} of a systematic approach to derive pure-state N-representability conditions for the 1RDM opened up new possibilities for functional development \cite{Theophilou2015, Benavides-Riveros2018}. The application of pure conditions narrows the variational space of the 1RDM, leading to improvements in energy. However, it's important to note that this doesn't enhance the reconstruction of the approximate functional itself. Let us reaffirm the fact that having a 1RDM representing a pure state does not guarantee the N-representability of the approximately reconstructed 2RDM, and as a result, this guarantee does not extend to the approximate functional either. 

For a long time, it was commonly believed that pure and ensemble universal functionals coincided within their mutual domain of pure state N-representable 1RDMs \cite{Valone1980, Nguyen-Dang1985}. However, recent work \cite{Schilling2018} revealed that the ensemble functional emerges as the lower convex envelope of the pure functional. Surprisingly, the pure functional even influences the behavior of the ensemble functional beyond its own domain of pure N-representable 1RDMs. On the other hand, applying pure N-representability conditions has proven to be prohibitively complex. Consequently, there has been a strong motivation to transition from pure 1RDMFT to ensemble 1RDMFT to mitigate the intricacies associated with the pure conditions. Fortunately, the concurrence of the pure and ensemble energy 1RDM functionals on the set of v-representable 1RDMs was recently confirmed \cite{Gritsenko2019}. This observation provides a rationale for the existing practice, in which only ensemble constraints are taken into account during the energy minimization process to ensure N-representability conditions. Therefore, our discussion is limited to physical electronic systems in their ground states, wherein the pure and ensemble functionals are indistinguishable.

As previously mentioned, the functionals currently in use rely on diagonal 1RDMs determined by the occupation numbers (ONs) of the associated NOs. Given the implicit dependence of approximate functionals on the 2RDM, it is more accurate to designate them solely as NOFs rather than categorizing them as pure 1RDM functionals. Ultimately, these functionals are only comprehended within the NO representation, including even the venerable functional derived from the Löwdin-Shull (LS) wavefunction \cite{Lowdin1955d} that accurately describes closed-shell two-electron systems. In this sense, NOFs can be viewed as approximate energy expressions that are, at best, derived from an approximate quantum state, specifically when adhering to N-representability conditions. An important consequence of this is that energy is not invariant under unitary transformations of the orbitals, which leads to the absence of a generalized Fockian when working with approximate functionals \cite{Piris2009a}. We can only demand this property from the exact functional.

There are two procedures to obtain a NOF: the top-down method and the bottom-up method \cite{Ludena2013, Piris2013e}. In the top-down approach, we begin by proposing an approximate N-particle wavefunction with the expansion coefficients explicitly expressed by the ONs. Subsequently, the energy automatically reduces to a NOF. Conversely, the bottom-up approach involves the proposal of a reconstruction for the 2RDM using ONs, without referencing the N-particle state. The functional is then constructed by incrementally integrating established N-representability conditions \cite{Mazziotti2012} into the proposed 2RDM reconstruction.

The top-down methodology inherently ensures N-representability, but, with few exceptions, it leads to viable expressions. In most cases, the method results in energy expressions that require additional quantities beyond the NOs and their ONs, making them not true NOFs. One very illustrative example is the previously mentioned functional derived from the LS wavefunction \cite{Lowdin1955d}. For two-electron systems, the top-down method leads to an energy that depends not only on NOs and ONs but also on phase factors. It has been demonstrated that in the weak correlation regime near Hartree-Fock (HF), all phases can be taken as negative, except for one phase that should have the opposite sign, corresponding to the highest occupation value. However, in other correlation regimes, alternating signs are observed \cite{Sheng2013}. Having to choose among a large number of phase combinations is referred to as the ``phase dilemma" \cite{Pernal2004, Mitxelena2018a}. 

Surprisingly, utilizing the phases derived from the near-HF case yields highly accurate results for two-electron systems, even in regimes of strong electronic correlation. This behavior is closely associated with the fact that, in this Coulombic system, the phases never change sign by varying the external one-body potential \cite{Giesbertz2013}. It should be noted that even in this relatively simple case, there is currently no established method to express the energy in terms of the complete 1RDM. Consequently, it is impossible to fully define the functional in a representation other than that of NOs. The LS energy with predefined phases cannot be classified as an authentic 1RDM functional. A bona fide 1RDM functional must exhibit independence from phase factors.

The bottom-up approach to generate a NOF was introduced by the author \cite{Piris2003} employing the cumulant expansion of the 2RDM \cite{Mazziotti1998, Kutzelnigg1999}. In this work, certain necessary N-representability conditions were imposed on the cumulant matrix expressed in terms of ONs. The use of positivity conditions (2,2) \cite{Mazziotti2012}, also known as D, Q, and G conditions, became a part of NOFT with the emergence of the PNOF \cite{Piris2006}. Appropriate forms of the two-electron cumulant lead to different implementations known in the literature as PNOFi (i=1-7) \cite{Leiva2005, Piris2007a, Piris2010, Piris2010a, Piris2011, Piris2014c, Piris2017}. The fundamental features of these functionals, as well as their achievements up to the year 2019, have been previously analyzed in several review articles \cite{Piris2013b, Piris2014a, Mitxelena2019}. In this review, the focus will be on PNOFs grounded in electron pairing \citep{Piris2018e}, which have exhibited remarkable success in addressing non-dynamic electron correlation. These NOFs yield results that align well with accurate wavefunction-based methods, particularly for small systems, since they account also for a significant portion of dynamic electron correlation related to intrapair interactions. 

The missing dynamic electron correlation was initially incorporated by perturbative corrections \citep{Piris2013c, Piris2014b, Piris2018b}. The latest implementation called NOF-MP2 \citep{Piris2018b} adds second-order M\o ller-Plesset (MP2) corrections to a reference Slater determinant wavefunction formed with the NOs of an approximate NOF. This approach has demonstrated its capability to achieve reasonable agreements in dissociation processes \cite{Piris2018b, Quintero-Monsebaiz2021}, exhibiting performance comparable to the accurate Complete Active Space Second-Order Perturbation Theory (CAS-PT2) method in hydrogen abstraction reactions \citep{Lopez2019}. Additionally, it has proven to be highly reliable for investigating the mechanistic aspects of chemical reactions in elementary reactions involving transition metal compounds \citep{Mercero2021}. Newly, an innovative development \citep{Rodriguez-Mayorga2021} revealed that applying a canonicalization procedure to the NOs allows for the integration of virtually any many-body perturbation method with a NOF.

The incorporation of perturbative corrections enhances the accuracy of absolute energies compared to the reference NOF values. However, it does not contribute to the improvement of the quality of the reference NOs and ONs. Achieving fully optimized correlated NOs and ONs requires a comprehensive optimization process, but this approach becomes computationally prohibitive for perturbative methods. Thus, it is more practical to address the deficiency in dynamic correlation using a broader NOF than PNOF7. 

Recently \cite{Piris2021b}, a NOF was proposed for electronic systems with any spin value regardless of the external potential, that is, a global NOF (GNOF). The latter has expanded upon the achievements of PNOF by reaching a more balanced distribution between dynamic and static correlations, thereby enhancing or even eliminating delocalization errors observed in PNOF7 \cite{Lew-Yee2022a, Lew-Yee2023b}. It has been successfully applied to investigate diverse chemical systems, including hydrogen models in one, two, and three dimensions \cite{Mitxelena2022}, iron porphyrin multiplicity \cite{Lew-Yee2023a}, carbenes \cite{Lew-Yee2023c}, and all-metal aromaticity \citep{Mercero2023}. It is important to highlight that the agreement achieved by GNOF with accurate wavefunction-based methods extends beyond relative energies to include absolute energies. This adds confidence to the method's reliability.

At present, the scientific community has access to an open-source implementation of NOF-based methods, \href{http://github.com/DoNOF}{DoNOF}, along with the accompanying in-house programs PyNOF and DoNOF.jl in modern programming languages Python and Julia. The latter implementations also provide support for computing accelerators through graphics processing units (GPUs). The associated software \cite{Piris2021,Lew-Yee2021}, is designed to address the ground-state energy minimization problem of an electronic system in terms of NOs and their ONs. Among the capabilities of the code for an accurate description of spin multiplet states are geometry optimization, natural and canonical representations of molecular orbitals, computation of ionization potential and electric moments, perturbative corrections for estimating dynamic correlation, NOF-based ab initio molecular dynamics, and the calculation of excited states.

The chapter is structured as follows: Section \ref{sec:NOF} covers fundamental concepts and notations relevant to NOF approximations. Subsequently, Section \ref{sec:pairNOF} introduces electron-pairing-based NOFs, specifically PNOF5, PNOF7, and GNOF, corresponding to independent-pair, interacting-pair, and global models for electron interactions. Section \ref{sec:spin} extends NOFs to multiplets while conserving the total spin. In Section \ref{sec:single-point}, we detail the optimization procedures for single-point calculations, with a primary focus on the procedures implemented in the open-source code \href{http://github.com/DoNOF}{DoNOF}, developed in our research group in Donostia. The chapter continues with Sections \ref{sec:geo-opt} and \ref{sec:aimd}, dedicated to geometry optimization and ab initio molecular dynamics, respectively, closely linked to the availability of analytical gradients in NOFT. Finally, the extension of NOFs to excited states is presented in Section \ref{sec:exc-sta}.

\section{ \label{sec:NOF} From exact RDMFT to NOF Approximations }

We consider an N-electron system described by the nonrelativistic Hamiltonian
\begin{equation}
\hat{\mathcal{H}}_{el} = \sum\limits _{ik}H_{ki}\hat{a}_{k}^{\dagger}\hat{a}_{i}+\frac{1}{2}\sum\limits _{ijkl}\left\langle kl|ij\right\rangle \hat{a}_{k}^{\dagger}\hat{a}_{l}^{\dagger}\hat{a}_{j}\hat{a}_{i}\label{Ham}
\end{equation}
where $H_{ki}$ denote the matrix elements of the one-particle part of the Hamiltonian involving the kinetic energy and the potential energy operators, $\left\langle kl|ij\right\rangle $ are the two-particle interaction matrix elements, whereas $\hat{a}_{i}^{\dagger}$ and $\hat{a}_{i}$ are the familiar fermion creation and annihilation operators associated with the complete orthonormal spin-orbital set $\left\{ | \phi_{i} \rangle \right\} $,
\begin{equation}
 \langle \phi_{k} | \phi_{i} \rangle =\int d{\bf x}\phi_{k}^{\ast}\left({\bf x}\right)\phi_{i}\left({\bf x}\right)=\delta_{ki}\label{ortho}
\end{equation}
with an obvious meaning of the Kronecker delta $\delta_{ki}$. Here, ${\bf x\equiv}\left({\bf r,s}\right)$ stands for the combined spatial and spin coordinates, ${\bf r}$ and ${\bf s}$, respectively. Atomic units are used. 

The Hamiltonian being studied is independent of spin coordinates, which implies that a state with a total spin of $S$ forms a multiplet. In other words, it represents a mixed quantum state that encompasses all conceivable $S_{z}$ values. Within this framework, there exist $\left(2S+1\right)$ energy-degenerate eigenvectors denoted as $\left|SM\right\rangle$. Consequently, a mixed state is characterized by the following N-particle density matrix statistical operator:
\begin{equation}
\mathfrak{\hat{D}}={\displaystyle {\displaystyle {\textstyle {\displaystyle \sum_{M=-S}^{S}}}}\omega_{M}\left|SM\right\rangle \left\langle SM\right|}\label{DM}
\end{equation}

In Eq. (\ref{DM}), $\omega_{M}$ are positive real numbers that sum one, so that $\mathfrak{\hat{D}}$ corresponds to a weighted sum of all accessible pure states. For equiprobable pure states, we take $\omega_{M}=(2S+1)^{-1}$.

The expectation value of (\ref{Ham}) reads as
\begin{equation}
  E_{el}=\sum\limits_{ik} H_{ki} \Gamma_{ki} + \sum\limits_{ijkl} \left\langle kl|ij \right\rangle D_{kl,ij}
\label{Energy}
\end{equation}
where the 1RDM and 2RDM elements are
\begin{equation}
\begin{array}{c}
\Gamma_{ki}={\displaystyle {\textstyle {\displaystyle \sum_{M=-S}^{S}}}}\omega_{M}\left\langle SM\right|\hat{a}_{k}^{\dagger}\hat{a}_{i}\left|SM\right\rangle \\
D_{kl,ij}={\displaystyle {\textstyle {\displaystyle \frac{1}{2}\sum_{M=-S}^{S}}}}\omega_{M}\left\langle SM\right|\hat{a}_{k}^{\dagger}\hat{a}_{l}^{\dagger}\hat{a}_{j}\hat{a}_{i}\left|SM\right\rangle 
\end{array}
\end{equation}

The Löwdin normalization is employed, ensuring that the traces of the matrices $\Gamma$ and D correspond to the total number of electrons and electron pairs, respectively. Both matrices exhibit important properties \cite{Piris2007}: they are Hermitian, positive semidefinite, and bounded. When dealing with eigenstates of $\widehat{S}_{z}$, only density matrix blocks that preserve the total number of spins of each type remain non-zero. Specifically, within the 1RDM, there are two discernible non-zero blocks, namely $\varGamma^{\alpha\alpha}$ and $\varGamma^{\beta\beta}$. In contrast, the 2RDM exhibits three independent non-zero blocks, namely $D^{\alpha\alpha}$, $D^{\alpha\beta}$, and $D^{\beta\beta}$. Notably, the parallel-spin components of the 2RDM must adhere to antisymmetry, whereas $D^{\alpha\beta}$ lacks any particular symmetry requirement.

The one-matrix ${\Gamma}$ can be diagonalized by a unitary transformation of the spin-orbitals $\left\{ \phi_{i}\left({\bf x}\right)\right\} $ with the eigenvectors being the NOs and the eigenvalues $\left\{ n_{i}\right\} $ representing the ONs of the latter,
\begin{equation}
\Gamma_{ki}=n_{i}\delta_{ki}\label{1matrix}
\end{equation}

Restriction of the ONs to the range $0\leq n_{i}\leq1$ represents a necessary and sufficient condition for ensemble N-representability of the 1RDM \cite{Coleman1963}.

The final term in Eq. (\ref{Energy}) represents the electron-electron potential energy $V_{ee}$, which explicitly depends on the 2RDM. We will approximate the exact RDM functional (\ref{Energy}) using the following approach:

\begin{equation}
  E_{el}\left[\left\{ n_{i},\phi_{i}\right\} \right] = \sum\limits _{i}n_{i}H_{ii}+\sum\limits _{ijkl}D[n_{i},n_{j},n_{k},n_{l}]\left\langle kl|ij\right\rangle \label{ENOF}
\end{equation}
where $D[n_{i},n_{j},n_{k},n_{l}]$ represents the reconstructed 2RDM from the ONs. We neglect any explicit dependence of D on the NOs themselves given that the energy functional has already a strong dependence on the NOs via the one- and two-electron integrals. In this context, it is worth noting that our NOs are the ones that diagonalize $\Gamma$ associated with our approximate NOF (\ref{ENOF}).

Given the persistent dependence of an approximate functional on D \cite{Donnelly1979}, the resulting energy lacks invariance under unitary transformations of the orbitals. Consequently, this prevents the existence of the corresponding extended Fockian matrix for energy minimization through direct diagonalization. This dependence also gives rise to the functional N-representability issue, which concerns the conditions necessary for ensuring a one-to-one correspondence between $E_{el}[\mathfrak{D}] \equiv E_{el}[\Gamma,\text{D}]$ and $E_{el}\left[\left\{ n_{i},\phi_{i}\right\} \right]$. This matter is clearly related to the N-representability of the reconstructed 2RDM.

Due to the complexity of the necessary and sufficient conditions for ensuring that D corresponds to an N-particle $\mathfrak{D}$, any approximation for the energy functional must, at the very least, satisfy manageable necessary conditions for the N-representability of the two-matrix. The well-known (2,2)-positivity conditions are a familiar example. These conditions dictate that the two-electron density matrix (D), the electron-hole density matrix (G), and the two-hole density matrix (Q) must be positive semidefinite. It is precisely these conditions that our functionals meet \cite{Piris2010a, Piris2013b, Piris2018d}.

In general, the 2RDM depends on four indices, making it computationally expensive. We employ a two-index reconstruction based on the cumulant expansion of the 2RDM \cite{Mazziotti1998, Kutzelnigg1999} within the spin-restricted formulation. This specific reconstruction \cite{Piris2006} involves the introduction of two auxiliary matrices, $\Delta[n_{i},n_{j}]$ and $\Pi[n_{i},n_{j}]$. The (2,2)-positivity conditions impose strict inequalities on the off-diagonal elements of the $\Delta$ and $\Pi$ matrices, whereas the conservation of the total spin allowed to derive the diagonal elements \cite{Piris2009}. Different forms of these matrices have led to the development of the JKL-only family of functionals, denoted as PNOFi (i=1-7) \cite{Leiva2005, Piris2007a, Piris2010, Piris2010a, Piris2011, Piris2014c, Piris2017}. Here, J and K represent the standard Coulomb and exchange integrals, respectively, while L corresponds to the exchange-time-inversion integral \cite{Piris1999}.

Determining the sign of $\Pi[n_{i},n_{j}]$ presents a challenge, as there is no straightforward method to ascertain it. Consequently, numerous potential sign combinations for terms containing $\Pi$ arise, which is commonly referred to as the phase dilemma \cite{Mitxelena2018a}. In the simplest scenario involving two electrons, an accurate NOF has already been established through the exact LS wavefunction \cite{Lowdin1955d}. This achievement serves as a compelling rationale for adopting electron pairs as fundamental units in NOF approximations.

\section{ \label{sec:pairNOF} Electron-Pairing-Based NOFs }

Let us consider $\mathrm{N_{I}}$ unpaired electrons, which determine the system's total spin $S$. The remaining electrons, $\mathrm{N_{II}} = \mathrm{N-N_{I}}$, form pairs with opposite spins, resulting in a net spin of zero for the $\mathrm{N_{II}}$ electrons combined. In the absence of unpaired electrons ($\mathrm{N_{I}}=0$), the energy functional naturally simplifies to a NOF that describes singlet states.

We focus on the mixed state of highest multiplicity: $2S+1=\mathrm{N_{I}}+1$ \citep{Piris2019}. For the ensemble of pure states $\left\{ \left| SM \right\rangle \right\}$, it is essential to note that the expectation value of $\hat{S}_{z}$ is zero. Consequently, the spin-restricted theory can be employed, even in cases where $S$ is non-zero. Consequently, all spatial orbitals are doubly occupied within the ensemble, ensuring equal occupancies for particles with $\alpha$ and $\beta$ spins ($\varphi_{p}^{\alpha} \left(\mathbf{r}\right) = \varphi_{p}^{\beta} \left(\mathbf{r}\right) = \varphi_{p} \left(\mathbf{r} \right), n_{p}^{\alpha}=n_{p}^{\beta}=n_{p}$).

In line with $\mathrm{N_{I}}$ and $\mathrm{N_{II}}$, we divide the orbital space $\Omega$ into two subspaces: $\Omega = \Omega_{\mathrm{I}} \oplus \Omega_{\mathrm{II}}$. $\Omega_{\mathrm{II}}$ is composed of $\mathrm{N_{II}}/2$ mutually disjoint subspaces $\Omega{}_{g}$. Each of which contains one orbital $\left|g\right\rangle $ with $g\leq\mathrm{N_{II}}/2$, and $\mathrm{N}_{g}$ orbitals $\left|p\right\rangle $ with $p>\mathrm{N_{II}}/2$, namely,
\begin{equation} 
\Omega_{g}=\left\{ \left|g\right\rangle ,\left|p_{1}\right\rangle ,\left|p_{2}\right\rangle ,...,\left|p_{\mathrm{N}_{g}}\right\rangle \right\} .\label{OmegaG}
\end{equation}

Taking into account the spin, the total occupancy for a given subspace $\Omega_{g}$ is 2, which is reflected in the following sum rule:
\begin{equation}
\sum_{p\in\Omega_{g}}n_{p}=n_{g}+\sum_{i=1}^{\mathrm{N}_{g}}n_{p_{i}}=1,\quad g=1,2,...,\frac{\mathrm{N_{II}}}{2}.\label{sum1}
\end{equation}

In general, $\mathrm{N}_{g}$ can be different for each subspace as long as it describes the electron pair well. For convenience, we usually take it the same for all subspaces $\Omega_{g}\in\Omega_{\mathrm{II}}$. The maximum possible value of $\mathrm{N}_{g}$ is determined by the basis set used in calculations. It is essential to note that orbitals within each subspace $\Omega_{g}$ undergo changes throughout the optimization process to find the most favorable orbital interactions. As a result, the orbitals are not static during the optimization process; they adapt to the specific problem.

From (\ref{sum1}), it follows that
\begin{equation}
2\sum_{p\in\Omega_{\mathrm{II}}}n_{p}=2\sum_{g=1}^{\mathrm{N_{II}}/2}\left(n_{g}+\sum_{i=1}^{\mathrm{N}_{g}}n_{p_{i}}\right)=\mathrm{N_{II}}.\label{sumNpII}
\end{equation}

\begin{figure}[ht]
\begin{centering}
\caption{\label{fig1} Illustrative example of splitting of the orbital space $\Omega$ into subspaces: $\Omega=\Omega_{\mathrm{I}}\oplus\Omega_{\mathrm{II}}=\Omega^{a}\oplus\Omega^{b},\,\Omega_{\mathrm{II}}=\Omega_{\mathrm{II}}^{a}\oplus\Omega_{\mathrm{II}}^{b}$. $\Omega^{a}$ ($\Omega^{b}$) denotes the subspace composed of orbitals above (below) the level $\mathrm{N}_{\Omega}=\mathrm{N_{II}}/2+\mathrm{N_{I}}$, that is, $\Omega^{a}\equiv p>\mathrm{N}_{\Omega}$ ($\Omega^{b}\equiv p\protect\leq\mathrm{N}_{\Omega}$). Similarly, $\Omega_{\mathrm{II}}^{b}\equiv p\protect\leq\mathrm{N_{II}}/2$ and $\Omega_{\mathrm{II}}^{a}\equiv p>\mathrm{N}_{\Omega}$. In this example, $S=1$ (triplet) and $\mathrm{N_{I}}=2$, so two orbitals make up the subspace $\Omega_{\mathrm{I}}$, whereas fourteen electrons ($\mathrm{N_{II}}=14$)
distributed in seven subspaces $\left\{ \Omega_{1},\Omega_{2},...,\Omega_{7}\right\} $ make up the subspace $\Omega_{\mathrm{II}}$. Note that $\mathrm{N}_{g}=2$ for all subspaces $\Omega{}_{g}\in\Omega_{\mathrm{II}}$, and $\mathrm{N}_{\Omega}=9$. The arrows depict the values of the ensemble occupancies, alpha ($\downarrow$) or beta ($\uparrow$), in each orbital.}
\par\end{centering}
\centering{}\includegraphics[scale=0.5]{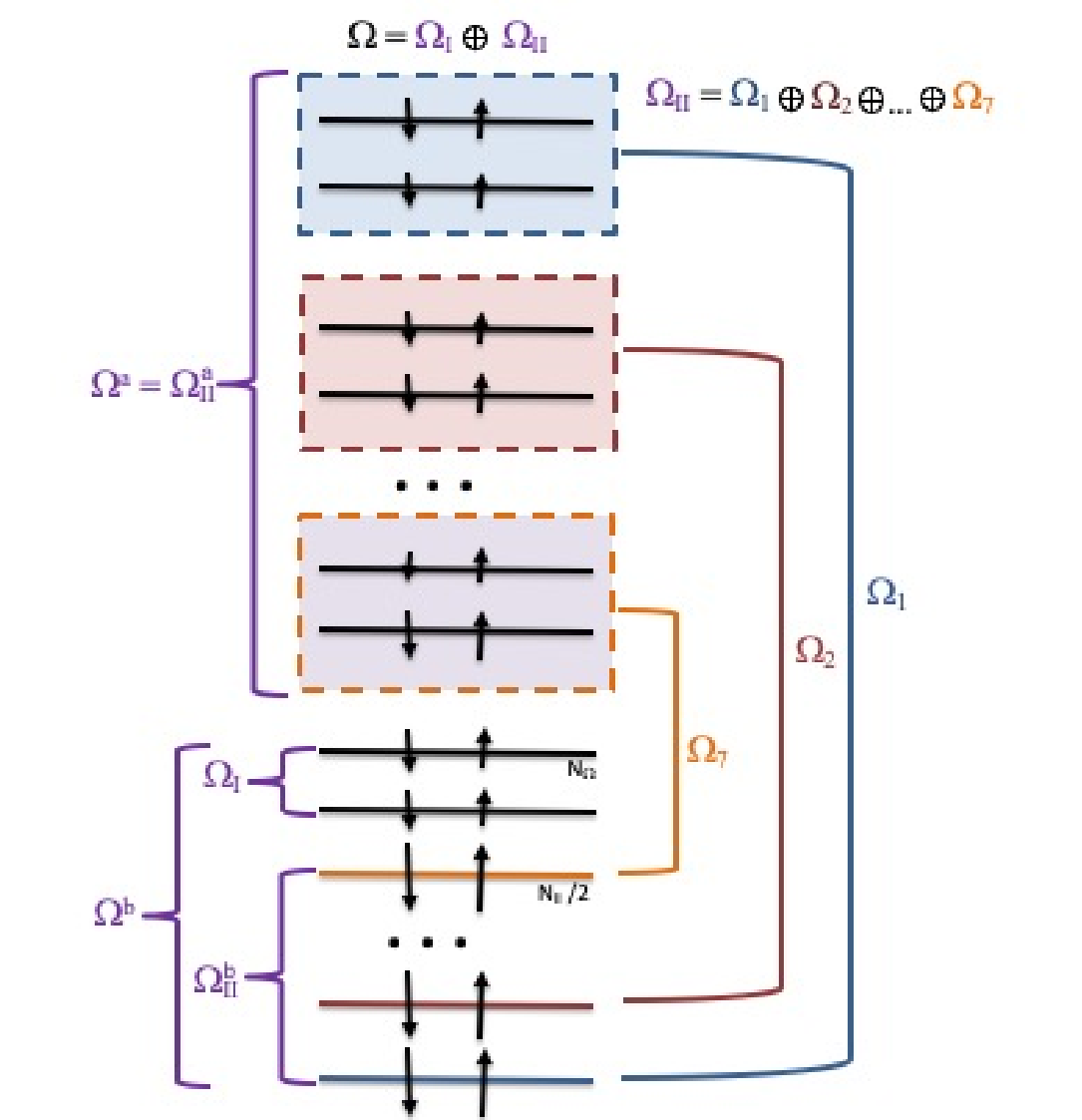}
\end{figure}

Similarly, $\Omega_{\mathrm{I}}$ is composed of $\mathrm{N_{I}}$ mutually disjoint subspaces $\Omega{}_{g}$. In contrast to $\Omega_{\mathrm{II}}$, each subspace $\Omega{}_{g} \in \Omega_{\mathrm{I}}$ contains only one orbital $g$ with $n_{g}=1/2$. It is important to emphasize that each orbital accommodates a single electron, but the specific spin state, whether $\alpha$ or $\beta$, is unknown. This leads to
\begin{equation}
2\sum_{p\in\Omega_{\mathrm{I}}}n_{p}=2\sum_{g=\mathrm{N_{II}}/2+1}^{\mathrm{N_{\Omega}}}n_{g}=\mathrm{N_{I}}.\label{sumNpI}
\end{equation}
In Eq. (\ref{sumNpI}), $\mathrm{\mathrm{N}_{\Omega}=}\mathrm{N_{II}}/2+\mathrm{N_{I}}$ denotes the total number of suspaces in $\Omega$. Taking into account Eqs. (\ref{sumNpII}) and (\ref{sumNpI}), the trace of the 1RDM is verified to be equal to the number of electrons: 
\begin{equation}
2\sum_{p\in\Omega}n_{p}=2\sum_{p\in\Omega_{\mathrm{II}}}n_{p}+2\sum_{p\in\Omega_{\mathrm{I}}}n_{p}=\mathrm{N_{II}}+\mathrm{N_{I}}=\mathrm{\mathrm{N}}.\label{norm}
\end{equation}

In Fig. \ref{fig1}, an illustrative example is shown. In this example, $S=1$ and $\mathrm{N_{I}}=2$, so two orbitals make up the subspace $\Omega_{\mathrm{I}}$, whereas fourteen electrons ($\mathrm{N_{II}}=14$) distributed in seven subspaces $\left\{ \Omega_{1},\Omega_{2},...,\Omega_{7}\right\} $ make
up the subspace $\Omega_{\mathrm{II}}$. The maximum value allowed by the basis set is $\mathrm{N}_{g}=2$.

\subsection{ \label{sec:PNOF5} Independent Pairs: PNOF5 }

In the realm of NOFs, the electron pairing approach made its debut with the inception of PNOF5 \cite{Piris2011, Piris2013e}. This approach incorporates the simplest method to meet the constraints imposed on the two-particle cumulant. Consequently, the matrix elements of D are segregated into contributions within and between subspaces. For intra-subspace blocks, only intrapair $\alpha\beta$ contributions emerge, specifically,
\begin{equation}
D_{pqrt}^{\alpha\beta\alpha\beta} = \frac{\Pi(n_p,n_r)}{2} \delta_{pq} \delta_{rt} \delta_{p\Omega_g} \delta_{r\Omega_g} \>,\quad g=1,2,...,\frac{\mathrm{N_{II}}}{2}
\label{intrag}
\end{equation}
where $\delta_{p\Omega_{g}}=1$ if $p\in\Omega_{g}$, or $\delta_{p\Omega_{g}}=0$ otherwise. 

The matrix elements $\Pi(n_p,n_r) = c(n_p)c(n_r)$, where $c(n_p)$ is defined by the square root of the ONs according to the following rule:
\begin{equation}
    c(n_p) = \left.
  \begin{cases}
    \phantom{+}\sqrt{n_p}, & p \leq \frac{\mathrm{N_{II}}}{2}\\
    -\sqrt{n_p}, & p > \frac{\mathrm{N_{II}}}{2} \\
  \end{cases}
  \right. \>\>, \quad p \in \Omega_{g} \in \Omega_{\mathrm{II}}
  \label{eq:PNOF5-roots}
\end{equation}
that is, the phase factor of $c_p$ is chosen to be $+1$ for the strongly occupied orbital of a given subspace $\Omega_g$, and $-1$ otherwise. It is important to emphasize that $D_{pp,pp}^{\alpha\beta\alpha\beta}=0$ for all $p\in\Omega_{\mathrm{I}}$. In fact, there can be no interactions between electrons with opposite spins in a singly occupied orbital, as each $\left|SM\right\rangle$ within an ensemble with $S \neq 0$ contains only one electron with either $\alpha$ or $\beta$ spin.

On the other hand, the contributions between different subspaces ($\Omega_g \neq \Omega_f$) are approximated in a manner resembling the HF method,
\begin{equation}
  D_{pqrt}^{\alpha\alpha\alpha\alpha} = \displaystyle \frac{n_p n_q}{2} \left( \delta_{pr} \delta_{qt} - \delta_{pt} \delta_{qr}\right) \delta_{p\Omega_f} \delta_{q\Omega_g}
\label{interaa}
\end{equation}
\begin{equation}
  D_{pqrt}^{\alpha\beta\alpha\beta} = \displaystyle \frac{n_p n_q}{2} \delta_{pr}\delta_{qt} \delta_{p\Omega_f}\delta_{q\Omega_g} \quad \quad \quad \quad \>\>
\label{interab5}
\end{equation}

We must remember that in the spin-restricted formalism we are using, the blocks $D^{\alpha\alpha}$ and $D^{\beta\beta}$ are equal. With this particular reconstruction of the 2RDM blocks, the energy (\ref{ENOF}) of PNOF5 can be succinctly expressed as
\begin{equation}
E_{el}\left[\mathrm{N},\left\{ n_{p},\varphi_{p}\right\} \right] = E^{intra}+E_{HF}^{inter} \label{pnof5}
\end{equation}

The intra-pair component is constructed by summing the energies $E_{g}$ of electron pairs with opposite spins and the single-electron energies of unpaired electrons, specifically,
\begin{equation}
E^{intra}=\sum\limits _{g=1}^{\mathrm{N_{II}}/2}E_{g}+{\displaystyle \sum_{g=\mathrm{N_{II}}/2+1}^{\mathrm{N}_{\Omega}}}H_{gg}
\label{Eintra}
\end{equation}
\begin{equation}
E_{g} = 2 \sum\limits _{p\in\Omega_{g}}n_{p}H_{pp} + \sum\limits _{q,p\in\Omega_{g}} \Pi(n_q,n_p) L_{pq}
\end{equation}
where $L_{pq}=\left\langle pp|qq\right\rangle$ are the exchange-time-inversion integrals \cite{Piris1999}. The inter-subspace HF term is 
\begin{equation}
\begin{array}{c}
E_{HF}^{inter} = \sum\limits _{f\neq g=1}^{\mathrm{N}_{\Omega}} \sum\limits _{p\in\Omega_{f}} \sum\limits _{q\in\Omega_{g}} n_{q}n_{p}\left(2J_{pq}-K_{pq}\right) = \sum\limits _{p,q=1}^{\mathrm{N}_{B}}{'\>} n_{q}n_{p}\left(2J_{pq}-K_{pq}\right) 
\end{array}
\label{Ehf}
\end{equation}
where $J_{pq}=\left\langle pq|pq\right\rangle$ and $K_{pq}=\left\langle pq|qp\right\rangle$ are the Coulomb and exchange integrals, respectively. $\mathrm{N}_{B}$ denotes the number of basic functions considered. The prime in the summation indicates that only the inter-subspace terms are taking into account ($p\in\Omega_f,q\in\Omega_g,f\neq g$).

To date, PNOF5 remains the sole NOF that has been derived through both top-down and bottom-up methodologies \cite{Piris2013e, Piris2018e}. The presence of a generating N-particle wavefunction attests to the pure-state N-representability of PNOF5 \cite{Pernal2013}.

Various performance assessments have consistently demonstrated that PNOF5 provides highly accurate descriptions of systems characterized by (nearly) degenerate one-particle states \cite{Piris2011, Matxain2011, Lopez2012, Matxain2013a, Piris2016b}. Notably, the outcomes achieved with PNOF5 concerning the electronic structure of transition metal complexes are particularly significant \cite{Ruiperez2013}. This functional adeptly captures the multiconfigurational aspects of the ground state of the chromium dimer, which is renowned as a benchmark molecule for quantum chemical methodologies due to its exceptionally complex electronic structure and potential energy curve.

PNOF5 has also demonstrated its efficacy in predicting the vertical ionization potentials and electron affinities of a carefully selected range of both organic and inorganic spin-compensated molecules, employing the extended Koopmans' theorem \cite{Piris2012}. The one-electron description provided by PNOF5 closely aligns with the orbitals derived from the Valence Bond (VB) method and those obtained through standard molecular orbital calculations \cite{Matxain2012, Matxain2012a, Matxain2013a}. In fact, an approximate NOF offers two distinct representations from the one-electron perspective, namely, the NO representation and the canonical orbital (CO) representation \cite{Piris2013}. Both sets of orbitals depict unique correlated one-electron scenarios, thereby complementing each other in the analysis of molecular electronic structure.

The property of size-consistency, coupled with the functional's ability to spatially localize NOs, positions PNOF5 as an exceptional choice for fragment-based computations. Notably, this approach has exhibited rapid convergence, enabling the efficient treatment of extended systems at a fraction of the overall computational cost \cite{Lopez2015a}.

The performance of PNOF5 was examined in describing the dissociation of small diatomic molecules \cite{Piris2013e}, where electron correlation primarily manifests as intra-pair interaction. It was observed that different values of the parameter $\mathrm{N}_{g}$ produced qualitatively accurate dissociation curves, with higher $\mathrm{N}_{g}$ values effectively capturing a larger portion of intra-pair correlation, as expected. It becomes evident that the primary limitation of PNOF5 lies in its inability to account for inter-pair electron correlation.

To address the absence of the correlation between pairs, two approaches were considered. One method involved employing a second-order size-consistent multiconfigurational perturbation theory, using PNOF5 generating wavefunction, resulting in the development of the PNOF5-PT2 method \cite{Piris2013c, Piris2014b}. The alternative approach aimed to incorporate the missing correlation from the outset by introducing interactions between electron pairs within the framework of NOFT, leading to the development of PNOF6 \cite{Piris2014c}. PNOF6 demonstrated superior treatment of both dynamic and non-dynamic electron correlations compared to PNOF5 \cite{Piris2015a, Lopez2015c, Piris2016, Mitxelena2016}. Notably, it was the only functional to exhibit consistent behavior when tested with exactly soluble models \cite{Cioslowski2015c}. Furthermore, PNOF6 effectively eliminates symmetry-breaking artifacts present in independent-pairs approaches when dealing with delocalized systems \cite{Ramos-Cordoba2015}. Nevertheless, it retrieves correlation energies that are lower in comparison to its predecessor.

\subsection{ \label{sec:PNOF7} Interacting Pairs: PNOF7}

To address inter-pair electron correlation and augment correlation energy compared to PNOF6, PNOF7 was introduced \cite{Piris2017}. The latter was enhanced through the judicious selection of sign factors for inter-pair interactions, resulting in a robust description of nondynamic correlation effects \cite{Mitxelena2018a}.

To derive PNOF7, the intra-subspace blocks, as represented by Eq. (\ref{intrag}), and the inter-subspace blocks for particles with parallel spins, denoted by Eq. (\ref{interaa}), are kept identical to those used in PNOF5. However, a new term is introduced, incorporating $\Phi_{p}=\sqrt{n_{p}h_{p}}$ with the hole $h_{p}=1-n_{p}$, for the inter-subspace blocks ($\Omega_g \neq \Omega_f$) for particles with opposite spins, namely,

\begin{equation}
\begin{array}{c}
   D_{pqrt}^{\alpha\beta\alpha\beta}=\left[{\displaystyle {\displaystyle \frac{n_{p}n_{q}}{2}\delta_{pr}\delta_{qt}-\frac{\delta_{f\Omega_{\mathrm{I}}}\delta_{g\Omega_{\mathrm{I}}}}{8}\delta_{pt}\delta_{qr}}}\right]\delta_{p\Omega_{f}}\delta_{q\Omega_{g}} \\
\\
   \qquad\quad - {\displaystyle \frac{\displaystyle\Phi_{p}\Phi_{r}}{2} \delta_{pq} \delta_{rt} \delta_{p\Omega_{f}} \delta_{r\Omega_{g}} (1-\delta_{f\Omega_{\mathrm{I}}}\delta_{g\Omega_{\mathrm{I}}})  }
\end{array}
\label{interab7}
\end{equation}

The resulting energy is
\begin{equation}
E_{el}\left[\mathrm{N},\left\{ n_{p},\varphi_{p}\right\} \right] = E^{intra}+E_{HF}^{inter}+\tilde{E}_{sta}^{inter} \label{pnof7}\\
\end{equation}
where the inter-subspace static component is written as 

\begin{equation}
\begin{array}{c}
   \tilde{E}_{sta}^{inter} = - \displaystyle { \sum\limits _{f\neq g=1}^{\mathrm{N}_{\Omega}} \sum\limits _{p\in\Omega_{f}} \sum\limits _{q\in\Omega_{g}} \Phi_{q}\Phi_{p}L_{pq} = - \sum\limits _{p,q=1}^{\mathrm{N}_{B}}{'\>}} \Phi_{q}\Phi_{p}L_{pq}
\end{array} \label{esta7}
\end{equation}

In Eq. (\ref{esta7}), it has been considered that $\Phi_{p}=1/2$ if $p\in\Omega_{\mathrm{I}}$, and $K_{pq}=L_{pq}$ for real spatial orbitals. The latter represents the typical choice in quantum chemistry. It becomes evident that the primary limitation of the approach (\ref{pnof7}) lies in the lack of inter-subspace dynamic electron correlation, as $\Phi_{p}$ exhibits notable values only when the ONs differ significantly from 1 and 0. Consequently, while PNOF7 can capture all intra-pair correlation like PNOF5, it is restricted to static inter-subspace correlation.

The effectiveness of PNOF7 has been verified in scenarios characterized by strong correlation. Indeed, we demonstrated the ability of PNOF7 to capture these correlation effects in challenging one-dimensional systems \cite{Mitxelena2020a} and two-dimensional systems \cite{Mitxelena2020b}, comparing them with calculations based on exact diagonalization, density matrix renormalization group, or quantum Monte Carlo methods.

As we are aware, the positivity conditions (2,2) imposed during the reconstruction of the PNOF 2RDM are necessary but not sufficient for its ensemble N-representability. Consequently, there may be situations where PNOFs could violate N-representability, leading to the so-called delocalization error. PNOF7 exhibited a small yet consistent charge delocalization error \cite{Lew-Yee2022a}, which has been associated with spurious contributions from static correlation due to the absence of dynamic interpair correlation terms in the functional.

\subsection{ \label{sec:GNOF} Global Natural Orbital Functional: GNOF }

As mentioned earlier, the absence of dynamic inter-subspace correlation can be addressed through perturbative corrections, yielding notable outcomes \cite{Piris2018b, Quintero-Monsebaiz2021, Lopez2019, Mercero2021, Rodriguez-Mayorga2021}. These post hoc corrections improve energy values over the reference PNOF7 results; however, they do not enhance the quality of the reference NOs and ONs. Full optimization remains the sole approach to achieve fully correlated ONs and NOs. Thus, it is advisable to address the absent dynamic correlation using a more general NOF than PNOF7. We refer to this functional as a global NOF (GNOF). It is worth noting that the term `global' is utilized instead of `universal' to distinguish our versatile approximate NOF for electronic systems, with any value of spin regardless of external potential, from Valone's exact counterpart. \cite{Valone1980}.

To derive GNOF, a novel reconstruction of the 2RDM was proposed \cite{Piris2021b}, taking into account the Pulay criterion for the division of electronic correlation into dynamic and nondynamic components based on ON values. This criterion sets a threshold of an occupancy deviation less than 0.01 from 1 or 0 for a NO to contribute to dynamic correlation, while larger deviations contribute to nondynamic correlation. It is evident that the PNOF7 functional form satisfies this criterion for static correlation, and thus, it is retained in GNOF. Additionally, a new term is introduced to account for dynamic correlation between subspaces. The new $\alpha\beta$ inter-subspace blocks ($\Omega_g \neq \Omega_f$) are defined as follows:
\begin{equation}
D_{pqrt}^{\alpha\beta\alpha\beta}=\left[{\displaystyle {\displaystyle \frac{n_{p}n_{q}}{2}\delta_{pr}\delta_{qt}-\frac{\delta_{p\Omega_{\mathrm{I}}}\delta_{q\Omega_{\mathrm{I}}}}{8}\delta_{pt}\delta_{qr}}}\right]\delta_{p\Omega_{f}}\delta_{q\Omega_{g}} + \> {\displaystyle \frac{{\displaystyle \Pi_{pr}^{d}} -{\displaystyle \Pi_{pr}^{s}}}{2}} \> \delta_{pq} \delta_{rt} \delta_{p\Omega_{f}} \delta_{r\Omega_{g}}
\label{interab}
\end{equation}

\begin{equation}
  \Pi_{pr}^{d} = \left[{ \Pi\left(n_{q}^{d},n_{p}^{d}\right) + n_{p}^{d}n_{r}^{d} }\right] \left( \delta_{p\Omega_{\mathrm{II}}^{b}} \delta_{r\Omega^{a}} + \delta_{p\Omega^{a}} \delta_{r\Omega_{\mathrm{II}}^{b}} + \delta_{p\Omega^{a}} \delta_{r\Omega^{a}} \right)
  \label{Pid}
\end{equation}

\begin{equation}
\Pi_{pr}^{s}=\Phi_{p}\Phi_{r}\left[{\delta_{p\Omega^{b}} \delta_{r\Omega^{a}} + \delta_{p\Omega^{a}}\delta_{r\Omega^{b}} + \delta_{p\Omega^{a}}\delta_{r\Omega^{a}}+\tfrac{1}{2}(\delta_{p\Omega_{\mathrm{II}}^{b}}\delta_{r\Omega_{\mathrm{I}}}+\delta_{p\Omega_{\mathrm{I}}}\delta_{r\Omega_{\mathrm{II}}^{b}})}\right]
\end{equation}
where $\Omega^{b}$ ($\Omega^{a}$) denotes the subspace composed of orbitals below (above) the level $\mathrm{N}_{\Omega}$, that is, $\Omega^{a}\equiv p>\mathrm{N}_{\Omega}$ ($\Omega^{b}\equiv p\protect\leq\mathrm{N}_{\Omega}$). On the other hand, $\Omega_{II}^{b}$ denotes the subspace composed of orbitals below the level $\mathrm{N_{II}}/2$ ($p\leq\mathrm{N_{II}}/2$), so interactions between two orbitals belonging to $\Omega_{II}^{b}$ are not considered in $\Pi^{d}$ and $\Pi^{s}$ matrices. In Eq. (\ref{Pid}), we can observe that $\Pi^d$ depends on the dynamic ONs, defined as:
\begin{equation}
n_{p}^{d}=n_{p}\cdot e^{-\left(\dfrac{h_{g}}{h_{c}}\right)^{2}},\,p\in\Omega_{g}\,,\enskip g=1,2,...,{\displaystyle \mathrm{N_{II}}/2}
\label{dyn-on}
\end{equation}

The value of $h_{c}$ in Eq. (\ref{dyn-on}) is fixed at $0.02\sqrt{2}$, indicating that the maximum value of $n_{p}^{d}$ is approximately 0.01, adhering to Pulay's criterion. It is important to note that $n_{p}^{d}$ does not account for the dynamic correlation of single electrons ($p\in\Omega_{\mathrm{I}}$). In Eq. (\ref{dyn-on}), a Gaussian function has been adopted to define the dynamic ON. Recent studies \cite{Johnson2023} in the context of Richardson-Gaudin states suggest that a simple exponential decay might yield more satisfactory results in regimes of intermediate correlation, i.e., where static or dynamic correlation does not predominantly prevail.

Considering real spatial orbitals and $n_{p}\approx n_{p}^{d}$, it is not difficult to verify that terms proportional to the product of the ONs in $\Pi^d$ will cancel out with the corresponding terms in $D^{\alpha\alpha}$ from Eq. (\ref{interaa}). As a result, only terms proportional to $\Pi\left(n_{q}^{d},n_{p}^{d}\right)$ will contribute to the energy. This functional form of inter-subspace $D^{\alpha\beta}$ blocks, when involved ONs deviate only slightly from 0 and 1, aligns with the functional form of intra-subspace $D^{\alpha\beta}$ blocks given by Eq. (\ref{intrag}).

Substituting in Eq. (\ref{ENOF}) the expressions (\ref{intrag}), (\ref{interaa}), and (\ref{interab}) for the 2RDM blocks, the GNOF is obtained:
\begin{equation}
E_{el}\left[\mathrm{N},\left\{ n_{p},\varphi_{p}\right\} \right] = E^{intra}+E_{HF}^{inter}+E_{sta}^{inter}+E_{dyn}^{inter} \label{gnof}
\end{equation}
where $E^{intra}$ and $E_{HF}^{inter}$ are given by Eqs. (\ref{Eintra}) and (\ref{Ehf}), respectively. The inter-subspace static component is expressed as
\begin{equation}
\begin{array}{c}
E_{sta}^{inter}=-\left({\displaystyle \sum_{p=1}^{\mathrm{N}_{\Omega}}\sum_{q=\mathrm{N}_{\Omega}+1}^{\mathrm{N}_{B}}+\sum_{p=\mathrm{N}_{\Omega}+1}^{\mathrm{N}_{B}}\sum_{q=1}^{\mathrm{N}_{\Omega}}}\right.\left.{\displaystyle +\sum_{p,q=\mathrm{N}_{\Omega}+1}^{\mathrm{N}_{B}}}\right)'
\Phi_{q}\Phi_{p}L_{pq} \\ \\
-\:\dfrac{1}{2}\left({\displaystyle \sum\limits _{p=1}^{\mathrm{N_{II}}/2}\sum_{q=\mathrm{N_{II}}/2+1}^{\mathrm{N}_{\Omega}}+\sum_{p=\mathrm{N_{II}}/2+1}^{\mathrm{N}_{\Omega}}\sum\limits _{q=1}^{\mathrm{N_{II}}/2}}\right)'
\Phi_{q}\Phi_{p}L_{pq}{\displaystyle \:-\: \dfrac{1} {4} \sum_{p,q = \mathrm{N_{II}}/2+1}^{\mathrm{N}_{\Omega}}} K_{pq}
\end{array} \label{esta}
\end{equation}
\\
\noindent whereas the inter-subspace dynamic energy is given by 
\begin{equation}
  E_{dyn}^{inter} = \left({\displaystyle \sum_{p=1}^{\mathrm{N_{II}}/2} \sum_{q=\mathrm{N}_{\Omega}+1}^{\mathrm{N}_{B}} + \sum_{p=\mathrm{N}_{\Omega}+1}^{\mathrm{N}_{B}} \sum_{q=1}^{\mathrm{N_{II}}/2} + \sum_{p,q=\mathrm{N}_{\Omega}+1}^{\mathrm{N}_{B}} }\right)' \left[ \Pi\left(n_{q}^{d},n_{p}^{d}\right) + n_{q}^{d}n_{p}^{d} \right] L_{pq}
\label{edyn}
\end{equation}

The functional (\ref{gnof}) has the ability to recover the entire intrapair electron correlation and incorporates interaction terms between orbitals that constitute both the pairs and the single electrons. The inter-subspace correlation, in turn, is composed of the sum of HF, static and dynamic terms. Its effectiveness in addressing strong correlation was assessed \cite{Mitxelena2022} in model hydrogen systems with different dimensionalities and electronic structures, including a 1D chain, a 2D ring, a 2D sheet, and a 3D compact pyramid. Additionally, two models representing strongly correlated Mott insulators-a 1D H50 chain and a 4×4×4 3D H cube-were investigated. The findings demonstrated that GNOF effectively handles both strong and weak correlation in a more balanced manner than its predecessor PNOFs.

In a previous study \cite{Lew-Yee2022a}, it was noted that the strictly N-representable PNOF5 tends to favor localized solutions, while PNOF7 may encounter the charge delocalization error in specific situations related to N-representability violations. Subsequent investigations \cite{Lew-Yee2023b} involved various analyses on GNOF, including assessments of charge distribution in super-systems consisting of two fragments, the stability of ionization potentials with an increase in system size, and potential energy curves for neutral and charged diatomic systems. GNOF was observed to effectively eliminate the charge delocalization error in numerous studied systems, or significantly improve the results compared to those obtained with PNOF7.

Recently \cite{Mercero2023}, GNOF was used to address the electron delocalization features of all-metal aromatic compounds, including the Al$_3$ ring-like cluster anion in its lowest-lying electronic states of different spin. The aromaticity was characterized by the multicenter index (MCI) and its $\pi$-fraction (MCI$_\pi$). The GNOF results turned out to be in very good agreement with the reference values obtained for benzene and cyclobutadiene with highly accurate correlated wavefunctions. The study revealed that GNOF accurately captures the multiple-fold aromaticity, both $\pi$- and $\sigma$-contributions, in the Al$_3^{-}$ states.

\section{\label{sec:spin} Conservation of the Total Spin S}

The nonrelativistic Hamiltonian (\ref{Ham}) commutes with the Hermitian spin operators, specifically with the total spin $\hat{S}^2$ of the system and one of its components, typically chosen as $\hat{S}_z$ (the spatial direction of quantization is irrelevant). This invariance of $\hat{\mathcal{H}}$ under spin rotations implies that its eigenvectors also serve as eigenvectors for the spin operators, and the corresponding quantum numbers $S$ and $M$ can consistently be identified as good quantum numbers.

Approximate functionals are not obligated to exhibit all the symmetries or, equivalently, possess the same good quantum numbers observed in exact 1RDMFT. In practice, relaxing the constraint on the conservation of total spin often ensures size-consistency and facilitates accurate energy predictions for molecular dissociations, among other scenarios. In these spin-unrestricted formulations, we allow symmetry breaking under $\hat{S}^2$ but not under $\hat{S}_z$. In other words, we fix the number of spin-up and spin-down electrons while allowing for spin contamination. These spin-unrestricted approaches are frequently employed in quantum chemistry, and, of course, energy reduction is achievable by breaking spin symmetry, albeit with a non-physical interpretation.

NOFs aiming to reproduce the expectation values of spin operators have been reported \cite{Leiva2007, Piris2009, Quintero-Monsebaiz2019}; however, these attempts have been limited to the high-spin component of the multiplet. In fact, there has even been speculation \cite{Cioslowski2020} that accurately describing spin-polarized systems is impossible with a reconstruction based on two indices. Nevertheless, through a reconstruction of the 2RDM for the whole multiplet, this becomes entirely feasible \citep{Piris2019}.

The ground state of a many-electron system with a total spin $S$ constitutes a multiplet given by Eq. (\ref{DM}). In this context, we direct our attention to the mixed-spin state of highest multiplicity ($2S+1=\mathrm{N_{I}}+1$), which is the sole nondegenerate state belonging to the quantum number $S=\mathrm{N_{I}}/2$ \cite{Paunez2000}. An essential aspect of our formulation is that the average spin projection within the entire ensemble is zero. This characteristic enabled us to employ a spin-restricted theory in the preceding section and put forth various approximations for the 2RDM blocks across the complete ensemble.

It is straightforward to confirm that the previous reconstructions for PNOF5, PNOF7, and GNOF lead to the conservation of the total spin $S$. In fact, the expectation value of the operator $\hat{S}^2$ is given by: 

\begin{equation}
  \mathrm{<}\hat{S}^{2}\mathrm{>}={\displaystyle \frac{\mathrm{N}\left(4-\mathrm{N}\right)}{4}}{\displaystyle +\sum\limits _{pq}}\left\{ D_{pqpq}^{\alpha\alpha\alpha\alpha}+D_{pqpq}^{\beta\beta\beta\beta}-2D_{pqqp}^{\alpha\beta\alpha\beta}\right\} 
\label{squareS}
\end{equation}

Proceeding with the following modification in the summations
\[
\sum\limits_{pq} \quad\rightarrow\quad {\displaystyle \sum_{g=1}^{\mathrm{N}_{\Omega}}\;}{\displaystyle \sum_{p,q\in\Omega_{g}}+\sum_{f\neq g=1}^{\mathrm{N}_{\Omega}}\;\sum_{p\in\Omega_{f}}\sum_{q\in\Omega_{g}}}
\]

\noindent and considering that the elements of the parallel-spin block of the 2RDM only have inter-subspace contributions, Eq. (\ref{interaa}), the trace of $D^{\sigma\sigma}$ with $\sigma=\alpha,\beta$ can be expressed as:
\begin{equation}
\sum\limits _{pq}D_{pqpq}^{\sigma\sigma\sigma\sigma}={\displaystyle \frac{1}{2}}{\displaystyle \sum_{f\neq g=1}^{\mathrm{N}_{\Omega}} \sum_{p\in\Omega_{f}} \sum_{q\in\Omega_{g}}}{\displaystyle n_{p}n_{q}} = {\displaystyle \frac{1}{2}\left\{ {\displaystyle \sum_{f\neq g=1}^{\mathrm{N_{II}}/2}} + \sum_{f=1}^{\mathrm{N_{II}}/2} \sum_{g=\mathrm{N_{II}}/2+1}^{\mathrm{N}_{\Omega}} + \sum_{f=\mathrm{N_{II}}/2+1}^{\mathrm{\mathrm{N}_{\Omega}}} \sum_{g=1}^{\mathrm{N_{II}}/2} \right.}
\nonumber
\end{equation}

\begin{equation}
{\displaystyle \left.+ \sum_{f\neq g=\mathrm{N_{II}}/2+1}^{\mathrm{N}_{\Omega}}\right\} }{\displaystyle \sum_{p\in\Omega_{f}}\sum_{q\in\Omega_{g}}} {\displaystyle n_{p}n_{q}} = {\displaystyle \frac{1}{2}\left\{ {\displaystyle \sum_{f\neq g=1}^{\mathrm{N_{II}}/2}}1 + \sum_{f=1}^{\mathrm{N_{II}}/2} \sum_{g=\mathrm{N_{II}}/2+1}^{\mathrm{N}_{\Omega}}\frac{1}{2} + \sum_{f=\mathrm{N_{II}}/2+1}^{\mathrm{\mathrm{N}_{\Omega}}} \sum_{g=1}^{\mathrm{N_{II}}/2}\frac{1}{2} \right.} 
\nonumber
\end{equation}

\begin{equation}
+ \left.{\displaystyle \sum_{f\neq g=\mathrm{N_{II}}/2+1}^{\mathrm{N}_{\Omega}}\frac{1}{4}}\right\} ={\displaystyle \frac{\mathrm{N_{II}\left(N_{II}-2\right)}}{8}+\frac{\mathrm{N_{II}N_{I}}}{4}+\frac{\mathrm{N_{I}\left(N_{I}-1\right)}}{8}}
\label{Traa}
\end{equation}

\noindent where we have considered Eq. (\ref{sum1}), and 
\begin{equation}
{\displaystyle \sum_{p\in\Omega_{g}}}n_{p}=n_{g}=1/2,\,\Omega{}_{g}\in\Omega{}_{\mathrm{I}}
\end{equation}

Taking into account the $\alpha\beta$ blocks of the 2RDM and noting the same contribution from the three considered functionals, the summation in the final term $\alpha\beta$ of Equation (\ref{squareS}) can be formulated as:

\begin{equation}
\sum\limits _{pq}D_{pqqp}^{\alpha\beta\alpha\beta}={\displaystyle \frac{1}{2}} \; {\displaystyle \sum_{g=1}^{\mathrm{N_{II}}/2}\;}{\displaystyle \sum_{p\in\Omega_{g}}}{\displaystyle n_{p}}-{\displaystyle \frac{1}{2}}{\displaystyle \sum_{f\neq g=\mathrm{N_{II}}/2+1}^{\mathrm{N}_{\Omega}}
\sum_{p\in\Omega_{f}}\sum_{q\in\Omega_{g}}}{\displaystyle n_{p}n_{q}}
\nonumber
\end{equation}

\begin{equation}
={\displaystyle \frac{1}{2}}\left\{ {\displaystyle {\displaystyle \sum_{g=1}^{\mathrm{N_{II}}/2}}\,1-\sum_{f\neq g=\mathrm{N_{II}}/2+1}^{\mathrm{N}_{\Omega}}\frac{1}{4}}\right\} ={\displaystyle \frac{\mathrm{N_{II}}}{4}}-{\displaystyle \frac{\mathrm{N_{I}\left(N_{I}-1\right)}}{8}}
\label{abba}
\end{equation}

By combining Eq. (\ref{squareS}) with Eqs. (\ref{Traa}) and (\ref{abba}), one obtains the ensemble average of the square of the total spin, expressed as:

\begin{equation}
 \mathrm{<}\hat{S}^{2}\mathrm{>} = {\displaystyle \frac{\mathrm{\left(N_{I}+N_{II}\right)}\left(4-\mathrm{N_{I}-N_{II}}\right)}{4}} + +2 \left[ {\displaystyle \frac{\mathrm{N_{II}\left(N_{II}-2\right)}}{8}+\frac{\mathrm{N_{II}N_{I}}}{4}} \qquad \qquad \right.
\nonumber
\end{equation}
\begin{equation}
{\displaystyle \left. + \; \frac{\mathrm{N_{I}\left(N_{I}-1\right)}}{8} \right]} {\displaystyle - \frac{\mathrm{N_{II}}}{2}} + {\displaystyle \frac{\mathrm{N_{I}\left(N_{I}-1\right)}}{4}} = {\displaystyle \frac{\mathrm{N_{I}}}{2}\left(\frac{\mathrm{N_{I}}}{2}+1\right) = S(S+1)}
\label{S2}
\end{equation}

This result aligns with a multiplet characterized by a total spin $S=\mathrm{N_{I}}/2$.

\section{ \label{sec:single-point} Single-Point Optimization Procedures}

Minimization of the energy functional, $E\left[\mathrm{N},\left\{n_{p},\varphi_{p}\right\} \right]$, is carried out subject to the orthonormality constraint for the real spatial orbitals:

\begin{equation}
\left\langle p|q\right\rangle =\int d\mathbf{r} \varphi_{p}\left(\mathbf{r}\right) \varphi_{q}\left(\mathbf{r}\right) = \delta_{pq}\label{orthos}
\end{equation}

\noindent whereas the occupancies adhere to the ensemble N-representability conditions $0\leq n_{p}\leq1$, and pairing sum rules (\ref{sum1}). The latter are crucial for orbitals within $\Omega_{\mathrm{II}}$ since $n_{p}=1/2$, $\forall p \in \Omega_{\mathrm{I}}$. This constitutes a constrained optimization problem, commonly resolved by independently optimizing the energy with respect to the ONs and NOs. The optimization with respect to all variables simultaneously, using typical algorithms that employ projected gradients, has proven to be inefficient for minimizing currently used NOFs \cite{Cances2008}. This inefficiency arises because, as we have pointed out, they are not proper functionals of the 1RDM.

Over the last two decades, several implementations, varying in effectiveness, have been developed \cite{Cohen2002, Herbert2003, Pernal2005, Giesbertz2010, Baldsiefen2013, Theophilou2015a, Wang2017, Yao2021, Wang2022, Lemke2022}. While optimization with respect to ONs has made significant strides and can now be conducted efficiently, optimizing with respect to NOs still requires substantial enhancements to render NOF-based methods competitive with density functional approximations (DFAs), currently the most efficient in quantum chemistry. Importantly, the orbital optimization cannot be simplified into a pseudo-eigenvalue problem. Moreover, it must be addressed in the molecular basis, not the atomic one, necessitating a four-index transformation to evaluate electron repulsion integrals at each iteration. Nevertheless, in numerous scenarios involving strong electronic correlation, NOFs are preferred over DFAs, and they can even compete with wavefunction-based methods due to their more favorable scaling with the  system's size. 

This section primarily delves into the algorithms implemented in the open-source code \href{http://github.com/DoNOF}{DoNOF} \cite{Piris2021,Lew-Yee2021}.

\subsection{Optimization of Occupation Numbers}

Constraints on $\left\{n_{p}\right\}$ can be automatically imposed by expressing the ONs through new auxiliary variables $\left\{\gamma_{p}\right\}$ and utilizing trigonometric functions. Similarly, the equality constraints (\ref{sum1}) can always be satisfied by leveraging the properties of these functions. Consequently, we transform the constrained minimization problem of the objective function with respect to ONs into the problem of minimizing energy concerning $\gamma$ auxiliary variables without restrictions on their values. This unconstrained optimization is computationally efficient and can be carried out using a successful technique like the conjugate gradient (CG) method, which has very modest storage requirements.

Let us define the ON of a strong occupied NO $\varphi_{g}$ as
\begin{equation}
n_{g}=\dfrac{1}{2}\left(1+cos^{2}\gamma_{g}\right)\>, \quad g=1,2,...,\mathrm{N_{II}}/2\label{soo}
\end{equation}
Considering the range $[0,1]$ for possible values of $cos^2\gamma_{g}$, it is evident that $1/2\leq n_{g}\leq1$. The pairing conditions (\ref{sum1}) can be expressed conveniently as
\begin{equation}
\sum_{i=1}^{\mathrm{N}_{g}}n_{p_{i}} = 1-n_{g} = h_{g}\>,\quad g=1,2,...,\mathrm{N_{II}}/2\label{sum1-h}
\end{equation}
where the hole $h_{g}$ in the orbital $\varphi_{g}$ is $h_{g}=\left(sin^{2}\gamma_{g}\right)/2$. The ONs of the rest of orbitals can be expressed through new auxiliary variables while satisfying Eq. (\ref{sum1-h}). Indeed, the ONs of each subspace $\Omega_{g}\in\Omega_{\mathrm{II}}$, Eq. (\ref{OmegaG}), can be set as
\begin{equation}
n_{p_{1}}=h_{g}sin^{2}\gamma_{p_{1}}, \quad n_{p_{2}}=h_{g}cos^{2}\gamma_{p_{1}}sin^{2}\gamma_{p_{2}}, 
\quad \cdots\\
\nonumber
\end{equation}
\begin{equation}
n_{p_{i}}=h_{g}cos^{2}\gamma_{p_{1}}cos^{2}\gamma_{p_{2}}\cdots cos^{2}\gamma_{p_{i-1}}sin^{2}\gamma_{p_{i}}, \quad \cdots \\
\nonumber
\end{equation}
\begin{equation}
n_{p_{\mathrm{N}_{g}-1}}=h_{g}cos^{2}\gamma_{p_{1}}cos^{2}\gamma_{p_{2}}\cdots cos^{2}\gamma_{p_{\mathrm{N}_{g}-2}}sin^{2}\gamma_{p_{\mathrm{N}_{g}-1}}\\
\nonumber
\end{equation}
\begin{equation}
n_{p_{\mathrm{N}_{g}}}=h_{g}cos^{2}\gamma_{p_{1}}cos^{2}\gamma_{p_{2}}\cdots cos^{2}\gamma_{p_{\mathrm{N}_{g}-2}}cos{}^{2}\gamma_{p_{\mathrm{N}_{g}-1}}\\
\label{woo}
\end{equation}

Verifying the equations in pairs from (\ref{woo}), starting with the last two and proceeding upwards using the resultant equation from each sum, it is evident that the constraint (\ref{sum1-h}) is consistently satisfied, thanks to the fundamental trigonometric identity. It is worth noting that by eliminating the equality constraints (\ref{sum1-h}) from the problem, we transition from having $\mathrm{N}_{g}$ unknown occupancies to $\mathrm{N}_{g}-1$ auxiliary $\gamma$-variables. Finally, considering that $sin^{2}\left(\gamma\right)$ and $cos^{2}\left(\gamma\right)$ always fall within the interval $[0,1]$ and that they multiply $h_{g}$, the remaining orbitals in $\Omega_{\mathrm{II}}$ are weakly occupied, i.e., $0\leq n_{p}\leq1/2$ if $p>\mathrm{N_{II}}/2\cap p\in\Omega_{\mathrm{II}}$.

\subsection{Optimization of Natural Orbitals}

For a fixed set of occupancies, the orthonormal conditions (\ref{orthos}) can be addressed using the Lagrange multipliers method. Introducing symmetric multipliers $\left\{\lambda_{qp}\right\}$ associated with the orthonormality constraints on the real spatial orbitals $\varphi_{p}$, the following auxiliary functional is defined:

\begin{equation}
\Omega=E_{el}-2{\displaystyle \sum_{pq}}\lambda_{qp}\left(\left\langle p|q\right\rangle -\delta_{pq}\right)\label{omega}
\end{equation}

The functional $\Omega$ must be stationary concerning variations in $\varphi_{p}$, meaning
\begin{equation}
\delta\Omega=\sum\limits _{p}\int d\mathbf{r}\delta\varphi_{p}\left(\mathbf{r}\right)\left[\frac{\delta E_{el}}{\delta\varphi_{p}\left(\mathbf{r}\right)}-{\displaystyle 4\sum_{q}}\lambda_{qp}\varphi_{q}\left(\mathbf{r}\right)\right]=0\label{variation}
\end{equation}

\noindent resulting in the following Euler equations
\begin{equation}
\frac{\delta E_{el}}{\delta\varphi_{p}\left(\mathbf{r}\right)}=4n_{p}\hat{H}\left(\mathbf{r}\right)\varphi_{p}\left(\mathbf{r}\right)+\frac{\delta V_{ee}}{\delta\varphi_{p}\left(\mathbf{r}\right)}={\displaystyle 4\sum_{q}}\lambda_{qp}\varphi_{q}\left(\mathbf{r}\right)
\label{lowdin}
\end{equation}

Irrespective of the chosen electron-pair-based NOF, the functional derivative of $V_{ee}$ with respect to $\varphi_{p}$ is contingent on the subspace $\Omega{}_{g}$ to which the orbital pertains. For orbitals within a subspace housing a single electron ($\varphi_{p}\in\Omega_{g}\in\Omega_{\mathrm{I}}$), only inter-subspace contributions emerge. In contrast, for orbitals within a subspace accommodating an electron pair ($\varphi_{p}\in\Omega_{g}\in\Omega_{\mathrm{II}}$), the intrapair contribution must also be considered.

The functional (\ref{omega}) must also be stationary concerning variations in Lagrange multipliers, leading to Eqs. (\ref{orthos}). Given a fixed set of ONs, we need to determine $\left\{\varphi_{p}\right\}$ and $\left\{\lambda_{qp}\right\}$ that satisfy both Eqs. (\ref{orthos}) and (\ref{lowdin}). This system of equations is nonlinear with respect to $\varphi_{p}$, posing a challenge for solution. By multiplying Eq. (\ref{lowdin}) with $\varphi_{p}$, integrating over $\mathbf{r}$, and taking into account Eq. (\ref{orthos}), we obtain
\begin{equation}
\lambda_{qp}=n_{p}H_{qp}+g_{pq} \>,\quad
g_{pq} = \frac{1}{4} \int d\mathbf{r}\frac{\delta V_{ee}}{\delta\varphi_{p}\left(\mathbf{r}\right)}\varphi_{q}\left(\mathbf{r}\right)
\label{lambda}
\end{equation}

Evaluating $\lambda_{qp}-\lambda_{pq}$, and noting that $H$ is a symmetric matrix, it straightforwardly follows that
\begin{equation}
\left(n_{p}-n_{q}\right)H_{qp}+g_{pq}-g_{qp}=0
\label{lamlam}
\end{equation}

Eq. (\ref{lamlam}) eliminates $\left\{\lambda_{qp}\right\}$ as variables in the problem; in other words, the original equations have been replaced by the system of Eqs. (\ref{orthos}) and (\ref{lamlam}) involving only the unknown $\left\{\varphi_{p}\right\}$. Introducing a set of known basis functions allows the integral differential equations to be transformed into algebraic ones; however, standard methods for solving the resulting nonlinear system converge very slowly.

Generally, the energy functional described in Eq. (\ref{ENOF}) does not maintain its invariance under an orthogonal transformation of the orbitals. Consequently, Eq. (\ref{lowdin}) cannot be simplified into a pseudo-eigenvalue problem by diagonalizing the $\mathbf{\lambda}$ matrix. However, at the extremum, the matrix of Lagrange multipliers must exhibit symmetry: $\lambda_{qp}=\lambda_{pq}$. This symmetrical property of $\mathbf{\lambda}$ can be leveraged to streamline the problem.

\subsubsection{Effective Potential}

The concept of an effective electron-electron potential $\upsilon$ for generating the NOs can be traced back to Gilbert's original paper \cite{Gilbert1975}. Considering real orbitals, Gilbert employed the formal identity:
\begin{equation}
\frac{\delta V_{ee}}{\delta\varphi_{p}\left(\mathbf{r}\right)} = 4n_{p} \varphi_{p}\left(\mathbf{r}\right) \widehat{\upsilon}\left(
\mathbf{r}\right)  
\label{gilbert}
\end{equation}
suggesting that $\Gamma$ and $\lambda$ can be simultaneously diagonalized by the same unitary transformation, implying that NOs would also be canonical orbitals. From Eq. (\ref{lowdin}) follows
\begin{equation}
n_{p}\left[\widehat{H}\left(\mathbf{r}\right)+\widehat{\upsilon}\left(\mathbf{r}\right)\right] \varphi_{p}\left(\mathbf{r}\right) = \lambda_{pp} \varphi_{p}\left(\mathbf{r}\right) 
\end{equation}

Ensuring the trace of $\Gamma$ is equal to N and taking $\upsilon_{pp}=\partial V_{ee} / \partial n_p$, it implies that NOs would exhibit an essentially degenerate eigenvalue spectrum ($\varepsilon_p = \lambda_{pp}/n_p = \mu $) \cite{Gilbert1975}. Regrettably, none of the currently known NOFs satisfies the formal relation (\ref{gilbert}).

Returning to Eq. (\ref{lamlam}), and defining a symmetric matrix $\upsilon$ with elements given by
\begin{equation}
\upsilon_{qp} = \frac {g_{pq}-g_{qp}}{n_{p}-n_{q}} \>,
\label{vqp}
\end{equation}
we can introduce a formal generalized Fockian ($\widehat{F}$) with matrix elements $F_{qp}=H_{qp}+\upsilon_{qp}$. The associated Hermitian operator $\widehat{\upsilon}$ can also serve as an optimal nonlocal electron-electron potential. Indeed, in accordance with Eq. (\ref{lamlam}), F satisfies the commutation relation:
\begin{equation}
\left[\mathrm{F},\Gamma\right] = \left[\mathrm{H}+\upsilon,\Gamma\right] = 0 
\label{conm}
\end{equation}
at the extremum. Therefore NOs could be obtained as solutions of the following eigenproblem: 
\begin{equation}
\widehat{F}\left(\mathbf{r}\right) \varphi_{p}\left(\mathbf{r}\right)
= \left[\widehat{H}\left(\mathbf{r}\right) + \widehat{\upsilon}\left(
\mathbf{r}\right)\right] \varphi_{p}\left(\mathbf{r}\right) = \varepsilon_{p} \varphi_{p}\left(\mathbf{r}\right) 
\label{eigen}
\end{equation}

It's essential to highlight that Eq. (\ref{vqp}) does not fully determine $\upsilon$. Specifically, the diagonal elements $\upsilon_{pp}$ and the off-diagonal elements $\upsilon_{qp}$ for orbitals with equal ONs ($n_q=n_p$) may take arbitrary values. Pernal proposed \cite{Pernal2005} that, even for NOFs implicitly dependent on the 1RDM, the non-local effective potential's kernel is the functional derivative of the energy functional with respect to the 1RDM, which implies that $\upsilon_{pp}=\partial V_{ee}/\partial n_p$. This transforms the challenge of finding optimal NOs into an iterative eigenproblem for F. Unfortunately, this iterative process is intrinsically divergent due to the eigenvalue degeneracy mentioned above \cite{Requist2008}. To tackle this issue, the frequently used level-shifting method can be employed \cite{Pernal2005, Requist2008}, although its successful application has been mainly demonstrated in the case of two electron systems.

\subsubsection{Iterative Diagonalization (ID) Method}

It is pertinent to ask whether, after all, minimizing a NOF can be posed as a self-consistent eigenvalue problem. The answer is affirmative \cite{Piris2009} and is once again based on leveraging the symmetry of the matrix $\lambda$.

Let us define the off-diagonal elements of a symmetric matrix $\mathcal{F}$ as
\begin{equation}
\mathcal{F}_{qp}=\theta\left(p-q\right)[\lambda_{qp}-\lambda_{pq}]+\theta\left(q-p\right)[\lambda_{pq}-\lambda_{qp}]
\label{Fockian}
\end{equation}

\noindent where $\theta\left(x\right)$ represents the unit-step Heaviside function. As per Eq. (\ref{lamlam}), $\mathcal{F}_{qp}$ becomes zero at the extremum, allowing matrices $\mathcal{F}$ and $\Gamma$ to be simultaneously diagonalized at the solution. Consequently, obtaining the $\varphi_{p}$'s that satisfy Eqs. (\ref{lamlam}) can be achieved through the ID of the matrix $\mathcal{F}$. The off-diagonal elements of matrix $\mathcal{F}$ and the effective Fockian F fundamentally differ by the factor of the occupancy difference. This distinction proves to be a significant advantage of $\mathcal{F}$ over F, as it circumvents convergence issues related to eigenvalue degeneracy. Moreover, $\mathcal{F}$ ensures the automatic satisfaction of (\ref{orthos}) due to the orthogonality inherent in its eigenfunctions.

Yet again, the determination of diagonal elements cannot be directly inferred from the symmetry property of $\lambda$; however, $\mathcal{F}_{pp}$ can be established with the aid of an aufbau principle \cite{Piris2009}. According to this principle, by considering an almost diagonal matrix $\mathcal{F}^{0}$ and choosing $\mathcal{F}_{qq}^{0}>\mathcal{F}_{pp}^{0}$ to make the first-order energy contribution negative, the energy is bound to decrease upon the diagonalization of $\mathcal{F}^{0}$. Consequently, the aufbau principle facilitates the definition of the diagonal elements.

In \href{http://github.com/DoNOF}{DoNOF}, the values from the previous diagonalization of $\mathcal{F}$ are preserved to be used as diagonal elements in each iteration. In the absence of static correlation, a suitable initial set for $\left\{\mathcal{F}_{pp}^{0}\right\}$ is the values obtained after energy optimization concerning $\gamma$ and a single diagonalization of the symmetrized $\lambda$-matrix, $\left(\lambda_{pq}+\lambda_{qp}\right)/2$, computed with the HF orbitals. However, if static correlation is significant, it is generally more effective to substitute the HF orbitals with the eigenvectors of H.

On the other hand, achieving a well-scaled $\mathcal{F}$ becomes decisive for energy decrease. In \href{http://github.com/DoNOF}{DoNOF} implementation, the matrix elements of $\mathcal{F}$ are appropriately scaled, ensuring that the value of each $\mathcal{F}_{pq}$ is of the same order of magnitude and remains below a specified upper bound $\zeta$.

It is well-known that iterative methods often suffer from slow convergence. To expedite convergence in \href{http://github.com/DoNOF}{DoNOF}, the DIIS extrapolation technique \cite{Pulay1980} is employed, utilizing an error vector in each iteration related to the gradient of the electronic energy with respect to $\varphi_{p}$. For this purpose, the off-diagonal elements of the $\mathcal{F}$ given by Eq. (\ref{Fockian}) are used to construct the error vector (bearing in mind that $\mathcal{F}_{pq} \rightarrow 0$ at the convergent solution).

It is important to note that the orbitals belonging to different subspaces vary throughout the optimization process until the most favorable orbital interactions are found, that is, there is no impediment to mixing orbitals from the different subspaces to arrive at the optimal orbitals. Consequently, the orbital optimization procedure is independent of the selected initial orbital coupling, although a proper initial guess favors a faster convergence of this procedure.

The scaling and DIIS techniques often ensure convergence in the procedure based on IDs of the $\mathcal{F}$ matrix. However, achieving highly accurate convergences becomes excessively slow due to a substantial prefactor in the cost scaling. It is clear that additional techniques are needed to accelerate convergence in the final steps of the method.

\subsubsection{Orbital rotations}

The combination of the ID method with the well-known orbital rotation technique \cite{Dalgaard1978, Shepard1987, Bozkaya2011, Elayan2022} can expedite the optimization of the NOs. From a computational perspective, the orbital rotation technique in NOFT exhibits similarities with post-HF methods traditionally applied to a few dozen orbitals. Nevertheless, recent advancements \cite{Kreplin2020} have extended its applicability to larger systems.

The diagonalization-free orbital optimization for NOF approximations can be traced back two decades \cite{Cohen2002, Herbert2003}. In this approach, the orthonormalized orbitals of an iteration, $\left\{\varphi_{p}\right\}$, undergo a relative rotation through an orthogonal transformation, given by
\begin{equation}
\tilde{\varphi}_{q} = \sum\limits_{p} U_{pq} \varphi_{p} = \sum\limits_{p} e^{k_{pq}} \varphi_{p}
\label{orbrot}
\end{equation}
where k is a skew matrix ($k_{qp}=-k_{pq}$) defining the independent orbital rotation parameters. Thus, the energy functional (\ref{ENOF}) transforms into a functional of the skew matrix, $E[\mathrm{k}]$, with its gradient and Hessian representing the first and second derivatives of the energy concerning the parameters $\left\{k_{pq}\right\}$, respectively,
\begin{equation}
g^\varphi_{pq} = \left. \frac{\partial E}{\partial k_{pq}}\right|_{k=0} \>, \qquad
H^\varphi_{pq,rs} = \left. \frac{\partial E}{\partial k_{pq}\partial k_{rs}} \right|_{k=0}
\end{equation}

General expressions for the above derivatives can be found in the relevant literature \cite{Bozkaya2011}. In our specific case, $g^\varphi_{pq}=\mathcal{F}_{pq}$ as given by Eq. (\ref{Fockian}), meaning that the NO gradient is defined by the symmetric matrix $\mathcal{F}$, which approaches zero at convergence.

The computational cost associated with the Hessian matrix $\mathrm{H}^\varphi$ is typically two orders of magnitude higher relative to the system's size compared to the gradient calculation. To circumvent the explicit construction and inversion of the molecular orbital Hessian, a CG algorithm for orbital rotations is implemented in \href{http://github.com/DoNOF}{DoNOF}.

There are two advantages to minimizing using orbital rotations compared to the ID method. The first is that, since k is a skew matrix, its diagonal is always zero. Therefore, the diagonal elements are not part of the parameters to be optimized. The second advantage is that, when using methods like the CG, we always have the assurance of decreasing the energy from one iteration to another, which is not guaranteed with the diagonalizations of $\mathcal{F}$.

As the system size grows, the ID method is more computationally efficient than optimization through orbital rotations. Considering the advantages of each method, employing the ID method initially and subsequent refinement using orbital rotations can expedite the optimization of the NOs. This is particularly beneficial when dealing with a substantial number of ONs with very small values, which tends to flatten the energy landscape during the minimization process.

\subsection{ \label{sec:efic} Computational Efficiency}

In current implementations, the optimization of the ONs and NOs alternates. We commence with an initial guess and engage in a loop embedding algorithm. Here, we optimize the ONs while keeping the NOs fixed during external iterations. Subsequently, we optimize the NOs while maintaining the ONs fixed in internal iterations, and this iterative process continues. 

We expand the NOs in a fixed atomic basis set, namely,
\begin{equation}
\varphi_{p}\left(\mathbf{r}\right)=\sum_{\upsilon=1}^{\mathrm{N}_{B}}\mathcal{C}_{\upsilon p}\zeta_{\upsilon}\left(\mathbf{\mathbf{r}}\right)\,,\quad p=1,\ldots,\mathrm{N}_{B}\label{lcao}
\end{equation}
Thus, bearing in mind that $L_{pq}=K_{pq}$ for real spatial orbitals, we calculate the requisite Coulomb and exchange integrals as follows:
\begin{equation}
J_{pq}=\sum_{\mu,\upsilon=1}^{\mathrm{N}_{B}}\Gamma_{\mu\upsilon}^{p} J_{\mu\upsilon}^{q}\, \quad K_{pq}=\sum_{\mu,\upsilon=1}^{\mathrm{N}_{B}}\Gamma_{\mu\upsilon}^{p} K_{\mu\upsilon}^{q}
\label{JKpq}
\end{equation}
where
\begin{equation}
\Gamma_{\mu\upsilon}^{p}=\mathcal{C}_{\mu p}\mathcal{C}_{\upsilon p}, \quad
J_{\mu\upsilon}^{q}=\sum_{\eta,\delta=1}^{\mathrm{N}_{B}}\Gamma_{\eta\delta}^{q}\left\langle \mu\eta|\upsilon\delta\right\rangle, \quad K_{\mu\upsilon}^{q}=\sum_{\eta,\delta=1}^{\mathrm{N}_{B}}\Gamma_{\eta\delta}^{q}\left\langle \mu\eta|\delta\upsilon\right\rangle \label{Gpmn}
\end{equation}

For the sake of computational efficiency, the one-electron and two-electron integrals in the atomic basis are computed once and stored in memory. This initial step primarily involves the arithmetic scaling of $\mathrm{N}_{B}^{4}$, dominated by the atomic orbital (AO) electron repulsion integrals (AO-ERIs). While this step, involving the evaluation and storage of AO-ERIs, does not significantly contribute to computational time due to its upfront execution, it represents the most memory-intensive stage with a memory scaling of $\mathrm{N}_{B}^{4}$

Energy minimization is carried out in the molecular basis, requiring a four-index transformation to convert AO-ERIs into molecular orbital (MO) electron repulsion integrals (MO-ERIs). Examining Eqs. (\ref{JKpq})-(\ref{Gpmn}), it becomes evident that the four-index transformation of the ERIs scales as $\mathrm{N}_{B}^{5}$, with dominance attributed to calculations involving $J_{\mu\upsilon}^{q}$ and $K_{\mu\upsilon}^{q}$. While this operation is performed once for fixed orbitals during occupancy optimization, it must be repeated every time the orbitals change to generate the matrix $\mathcal{F}$ during orbital optimization. This is the reason why optimizing the NOs represents the most time-consuming step in the energy minimization process.

The scaling factor $\mathrm{N}_{B}^{5}$ is relatively lower compared to other procedures, such as those based on configuration interaction and coupled cluster approaches. However, the four-index transformation remains computationally intensive. Consequently, various strategies have been implemented to enhance the efficiency of the optimization procedures. 

Firstly, the parallelization of specific sections of the code responsible for the four-index transformation has proven effective, leading to significant performance improvements. Secondly, for certain NOFs, exploiting the option to sum over NO indices separately helps alleviate computational costs \cite{Giesbertz2016, Mitxelena2019}. This strategy incorporates the integral-direct formalism to NOF calculations \cite{Lemke2022}, enabling the reformulation of central quantities, such as two-electron integrals, in an AO-based fashion. This allows for on-the-fly recalculation instead of storing them in memory or on disk. Moreover, an AO-based integral-direct formalism not only reduces the memory demands but also enables the exploitation of many already-existing efficient methods used in standard SCF schemes, such as integral screening or the continuous fast multipole method for the evaluation of Coulomb matrices.

Finally, to reduce arithmetic and memory scaling, the resolution of identity (RI) approximation, also known as density fitting, is employed \cite{Lew-Yee2021}. This approximation represents the product of basis functions as a linear combination of auxiliary basis sets, resulting in decreased arithmetic and memory scaling factors and producing easy-to-handle intermediate matrices. In particular, the use of the RI approximation in PNOF calculations decreases the arithmetic scaling factor of the four-index transformation from fifth order to fourth order. The details of the implementation of the RI approximation in \href{http://github.com/DoNOF}{DoNOF} can be found in the associated reference \cite{Lew-Yee2021}.

The RI approximation has undergone testing on cycloalkanes containing up to nine carbon atoms, as well as other molecules of broad interest such as oxazole, borazine, coumarin, cyanuric chloride, benzene, thiepine, and thieno[2,3-b]thiophene \cite{Lew-Yee2021}. A comparative analysis between PNOF7-RI and PNOF7 calculations revealed a substantial reduction in computational time while maintaining a consistent accuracy. In all instances, restarting from PNOF7-RI calculation converges to the PNOF7 energy within a maximum of two outer iterations, resulting in a PNOF7 equivalent outcome in a significantly reduced timeframe. This progress allowed the application of NOF approximations to larger molecular systems of general chemical interest, exemplified by the challenging triplet-quintet gap in the iron(II) porphyrin complex (FeP) \cite{Lew-Yee2023a}. In particular, it was shown that GNOF can efficiently handle a substantial number of electrons and orbitals, specifically, 186 electrons in 465 orbitals within the 37-atom FeP, thereby broadening the scope of multiconfigurational treatment. Furthermore, methods that incorporate significant dynamic correlation, such as NOF-MP2 and GNOF, yield accurate predictions for the quintet-triplet gap in FeP.

An important recent advancement that strengthens these strategies is their integration into the FermiONs++ program package, which runs on graphics processing units (GPUs) \cite{Lemke2022}. Such developments have made it possible to perform calculations on large systems of chemical interest with tens of atoms, hundreds of electrons, and thousands of basis functions. Examples include the 117-atom 2-carbamate taxol, a derivative of the anticancer drug paclitaxel, and the 168-atom valinomycin molecule, a potassium ionophore and potential antiviral agent against coronaviruses.

\section{ \label{sec:geo-opt} Geometry Optimization}

Geometry optimization, along with single-point energy calculations, constitutes one of the most widely employed procedures in electronic structure theory. Efficient geometry optimization relies on access to the analytical gradients with respect to nuclear motion. They play a pivotal role in identifying and characterizing critical points on the energy surface, including minima and saddle points. Additionally, these gradients are crucial for investigating high-resolution molecular spectroscopy and understanding geometry-dependent molecular properties.

The development of efficient computation for analytic energy gradients concerning nuclear motion is currently a focal point in RDM-based methods \cite{Mullinax2019}. In the context of NOFT, the general formulation of analytic gradients was introduced in Ref. \cite{Mitxelena2017}, enabling the efficient calculation of equilibrium geometries for molecules in singlet states. Importantly, it was demonstrated that analytic gradients can be obtained through simple evaluation, eliminating the need for linear response theory or involving iterative procedures. The extension to non-singlet states was subsequently proposed \cite{Mitxelena2020c} by considering the entire multiplet state. As a result, equilibrium geometries of molecular systems with any spin value can be easily obtained while preserving the total spin (see Section \ref{sec:spin}).

The derivative of the total energy ($E$) in the AO representation with respect to the coordinate $x$ of nucleus $A$ is expressed as follows:
\begin{equation}
  {\displaystyle \frac{dE}{dx_{A}}}
= {\displaystyle \frac{\partial E_{nuc}}{\partial x_{A}}}
+ {\displaystyle \sum_{\mu\upsilon}} \Gamma_{\mu\upsilon} \dfrac{\partial H_{\mu\upsilon}} {\partial x_{A}}
+ {\displaystyle \sum_{\mu\upsilon\eta\delta}} D_{\mu\eta\upsilon\delta}
   \dfrac{\partial\left\langle \mu\eta\upsilon\delta \right\rangle }{\partial x_{A}}
- {\displaystyle \sum_{\mu\upsilon}}\lambda_{\mu\upsilon} \dfrac{\partial\mathcal{S_{\mu\upsilon}}}{\partial x_{A}}
\label{NOF-analy-grads}
\end{equation}
Here, $\lambda$ is the matrix of Lagrange multipliers, and S represents the overlap matrix. In Eq. (\ref{NOF-analy-grads}), the first term corresponds to the derivative of the nuclear energy ($E_{nuc}$), the second term represents the negative Hellmann-Feynman force, and the third term contains the explicit derivatives of two-electron integrals. The last term, known as the density force, arises from the implicit dependence of the orbital coefficients  $\left\{ \mathcal{C}_{\upsilon p}\right\} $ on geometry. The implicit dependence of ONs on geometry does not contribute to analytic gradients since $E_{el}$ is stationary with respect to variations in all of the ONs \cite{Mitxelena2017}.

Equation (\ref{NOF-analy-grads}) provides a general expression for obtaining energy gradients for any NOF, requiring only the explicit reconstruction of D for specific approximations. This equation reveals that the computational bottleneck for NOF gradients lies in $D_{\mu\eta\upsilon\delta}$ and the derivatives of two-electron integrals, which formally scale as $N_{B}^{5}$. To address this, integral derivatives are computed on-the-fly, allowing for efficient application of Schwarz's screening and resulting in savings in storage and computation times. Moreover, as highlighted in Section \ref{sec:efic}, computational costs are reduced significantly by summing over NO indices separately. This separability strategy leads to a notable reduction in computation time for both PNOF5 and PNOF7 \cite{Mitxelena2019}, regardless of the number of orbitals considered in the calculation determined by $N_B$.

In the derivation described in Ref. \cite{Mitxelena2018b}, the second derivatives of a NOF energy are obtained by differentiating Eq. (\ref{NOF-analy-grads}) with respect to another coordinate $y$ associated with nucleus $B$. The first two terms in this Hessian matrix involve the explicit derivatives of the core Hamiltonian and two-electron integrals, respectively. The subsequent terms include contributions from the NOs and ONs concerning the nuclear perturbation. Unlike first-order energy derivatives, computing the analytic Hessian requires knowledge of NOs and ONs at the perturbed geometry, which can only be obtained through the solution of coupled-perturbed equations \cite{Mitxelena2018b}. Consequently, the computation of second-order energy derivatives is significantly more demanding than evaluating the total electronic energy or gradients, both in terms of storage capacity and computational times. To address these challenges, our implementations utilize a numerical approach, specifically the $6N_{a}$ point formula ($N_{a}$ denoting the number of atoms), to calculate the Hessian. In this context, analytical evaluation of the Hessian is not necessarily much more efficient than numerical differentiation of analytical gradients.

Second-order energy derivatives enable the calculation of harmonic vibrational frequencies. This involves obtaining a set of $3N_{a}$ eigenvectors corresponding to normal modes and $3N_{a}$ eigenvalues corresponding to the harmonic vibrational frequencies. However, six eigenvalues corresponding to overall translation and rotation are not exactly zero at a general point on the energy surface. Specifically, for displacements that are not rigorously orthogonal in the $3N_{a}$ dimensional vector space to the gradient vector, the potential is not quadratic, leading to the possibility of rotational and translational contaminant modes. Consequently, the Hessian is projected to restrict the displacements to be orthogonal to the $3N_{a}$ dimensional vectors corresponding to the rotations and translations of the system.

Our calculations \cite{Piris2021b, Mitxelena2022, Mercero2023} have shown that the equilibrium geometries and harmonic vibrational frequencies obtained with GNOF exhibit satisfactory agreement with accurate theoretical methods and available experimental data, despite a slight underestimation of equilibrium distances and a slight overestimation of frequencies.

\section{ \label{sec:aimd} \textit{Ab Initio} Molecular Dynamics}

The ability to calculate gradients analytically also makes NOF-based \textit{ab initio} molecular dynamics (AIMD) \cite{marx_hutter_2009} simulations feasible. The underlying concept involves calculating the forces acting on the nuclei through electronic structure calculations as the molecular dynamics trajectory unfolds. This approach eliminates the need to pre-determine interatomic interaction potentials, which can pose challenges in chemically complex systems where the bonding pattern qualitatively changes during dynamics. The price to pay is that the accessible correlation lengths and relaxation times are shorter than those in standard molecular dynamics, however, AIMD circumvents the dimensionality bottleneck associated with precomputing PES.

In scenarios where electrons respond almost instantaneously to nuclear motion, the BO approximation allows the decoupling of electronic and nuclear problems, defining a potential energy surface (PES). While the BO approximation is generally accurate for the gas-phase dynamics of molecules in their electronic ground state (GS), it may face challenges in other situations, such as dynamics starting in an electronically excited state. Here, strong couplings between two or more PES occur, necessitating a quantum treatment of both nuclei and electrons. Non-adiabatic molecular dynamics becomes the preferred method for modeling these processes. 

Currently, only the BO-AIMD based on an approximate NOF has been implemented \cite{Santamaria2023}. The total energy of a molecule is expressed as:
\begin{equation}
E = E_{nuc}+E_{el} \>, \quad 
E_{nuc} = \sum_{A<B} \frac {Z_{A}Z_{B}}{R_{AB}}
\end{equation}
where $Z_{A}$ represents the atomic number of nucleus $A$, and $R_{AB}$ is the distance between nuclei $A$ and $B$. The derivative of the total energy with respect to the coordinate $x$ of nucleus $A$ is given by Eq. (\ref{NOF-analy-grads}). All derivatives have an explicit dependence on the nuclear coordinate $x_{A}$, so the force acting on each nucleus A $(\textbf{F}_{A}=-\nabla_{A} E)$ can be obtained by a single static evaluation at each time step for the fixed nuclear positions at that instant. Consequently, we can calculate the trajectories of the nuclei according to the classical equations of motion, but taking into account the quantization of the reactants, a procedure known as quasiclassical trajectory (QCT) method. It is important to note that the QCT method does not account for tunneling effects, and therefore, it may yield inaccurate results near the threshold energy.

NOF-based QCT calculations can be performed using the new molecular dynamics module implemented in \href{http://github.com/DoNOF}{DoNOF}, which allows the calculation of nuclear trajectories by determining ``on the fly'' the forces using NOF gradients (\ref{NOF-analy-grads}). Beeman's algorithm \citep{Beeman1976} is used to numerically integrate Newton's equations of motion, whereas the initial conditions are obtained using a standard Monte Carlo sampling procedure \citep{Truhlar1979}.

Among the available functionals, GNOF is the NOF of choice for dynamics due to its excellent balance between static and dynamic correlation. It provides accurate total energy values while preserving spin, even for systems with highly multiconfigurational character. Additionally, GNOF correlates all electrons with all available orbitals for a given basis set, a feat currently unattainable for large systems with current wavefunction-based methods. Regarding dynamics, the NOs and ONs vary along the trajectory, adapting at each time step to the most favorable interactions of the corresponding nuclear configuration.

\begin{figure*}[htbp]
\centering
\caption{Time evolution of the two strongly occupied natural orbitals involved in the bond pattern change during the reaction N($^4$S) + H$_2$($^1\Sigma$) $\rightarrow$ NH($^3\Sigma$) + H($^2$S). \bigskip}
{\includegraphics[scale=0.35]{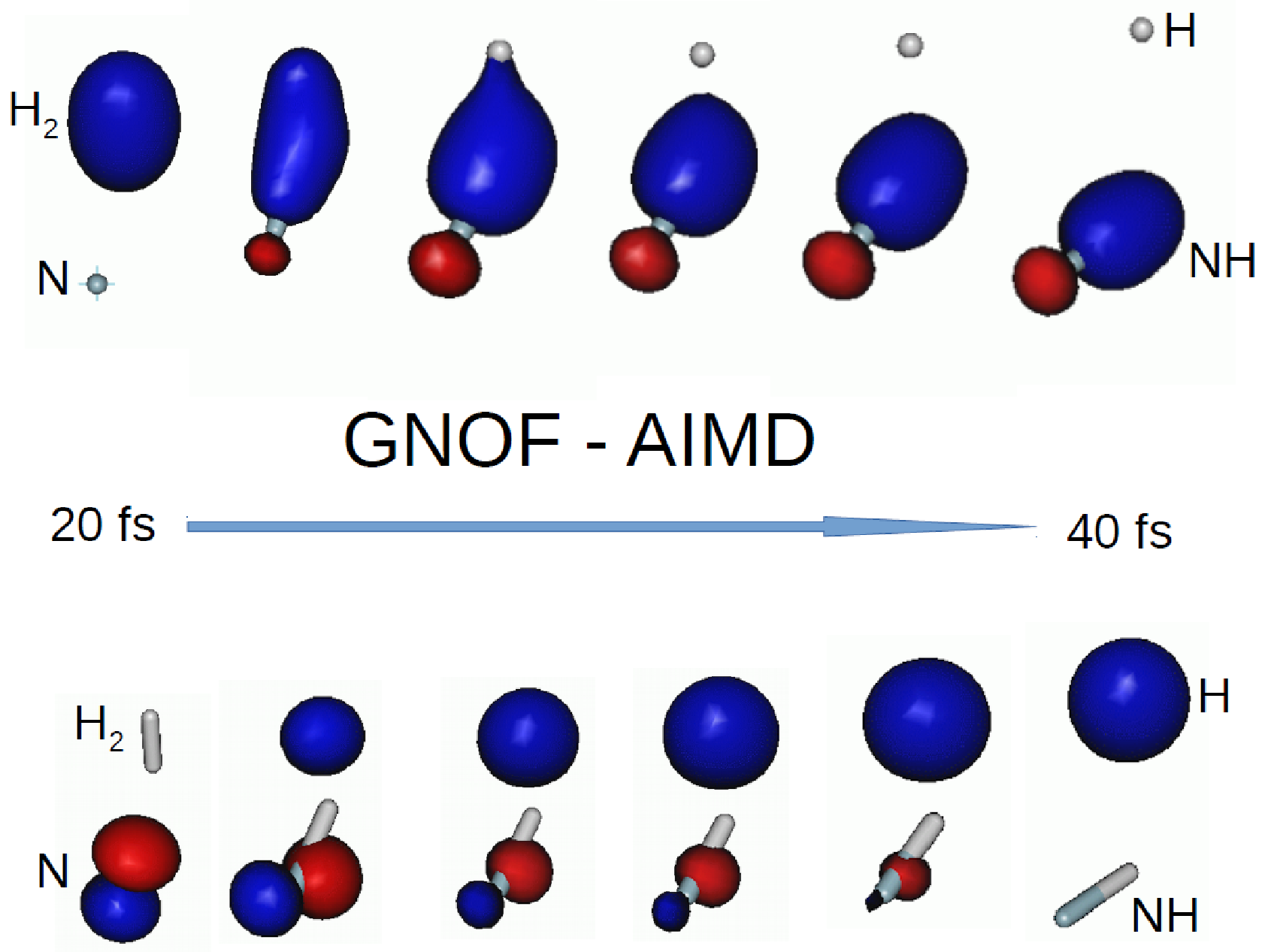}}
\label{orb-evol}
\end{figure*}

A significant advantage of GNOF-AIMD is the real-time observation of NO and their ON evolution during complex dynamics, providing detailed insights into the time-dependent electronic structure of such processes. In Ref. \cite{Santamaria2023}, the N($^4$S) + H$_2$($^1\Sigma$) $\rightarrow$ NH($^3\Sigma$) + H($^2$S) reaction was taken as a validation test. The reaction occurs via an abstraction mechanism dominated by the quartet ground state, and presents a forward experimental barrier of 1.4 $\pm$ 0.3 eV. The collision occurs mainly in the time range from 20 fs to 40 fs. Two NOs are responsible for the change of the bond pattern during the collision, whose  temporal evolutions are represented in Fig. \ref{orb-evol} by specific snapshots. We observe a NO that begins as a $\sigma$ ``ss'' bonding orbital of the H$_2$ singlet and transforms into the $\sigma$ ``sp'' bonding orbital of the NH triplet, while the other individually occupied NO transforms from a 2p atomic orbital of the N into the 1s atomic orbital of the H.

The impact of collision energy on integral cross-sections for distinct initial ro-vibrational states of H$_2$ and rotational-state distributions of NH products were also explored \cite{Santamaria2023}. The results demonstrated a strong concordance with prior high-quality theoretical findings. In conclusion, GNOF-AIMD introduces a promising avenue for research: AIMD grounded in natural orbital functionals.

\section{ \label{sec:exc-sta} Excited States}

In this segment, our focus lies in assessing the excitation energies of both charged ($E_{\nu}^{\mathrm{N}\mp1} - E_0^\mathrm{N}$) and neutral ($E_{\nu}^\mathrm{N} - E_0^\mathrm{N}$) states. Here, $E_{\nu}^\mathrm{M}$ represents the $\nu$-th eigenvalues of the M-electron system governed by the Hamiltonian (\ref{Ham}):
\begin{equation}
\hat{\mathcal{H}}_{el}\ket{\psi^\mathrm{M}_\nu} = E_{\nu}^\mathrm{M}\ket{\psi^\mathrm{M}_\nu}
\end{equation}
The calculations differ only in the excitation operator type used to generate a charged or neutral excited state. For charged excitations, we employ the extended Koopman's theorem \cite{Day1974, Smith1975, Day1975, Morrell1975}. Conversely, for neutral excitations, we use the Rowe's excitation operator \cite{Rowe1968}, leading to the equation-of-motion. In both cases, the coefficients defining these excited states are determined by the first and second-order RDMs of the ground state, as dictated by the fundamental equations of each method.

\subsection{Charge excitations: Extended Koopmans' Theorem (EKT)}

Within the domain of 1RDMFT, ionization potentials (IPs) and electron affinities (EAs) can be determined by calculating the energy differences between the relevant states. Nevertheless, unquestionably, the most straightforward approach involves employing the extension of Koopman's theorem to correlated states. Indeed, the EKT establishes a connection between the 1RDM and 2RDM of a Coulombic system and its IPs and EAs.

The equation for the EKT may be derived by expressing the wavefunction of the ($\mathrm{N}\mp1$)-electron system as the following linear combination
\begin{equation}
\ket{\psi_\nu^{\mathrm{N}-1}} 
= \hat{O}^\dagger_\nu \ket{\psi^{\mathrm{N}}_0}
= \sum\limits _{i}C_{i\nu}\hat{a}_{i}\ket{\psi^{\mathrm{N}}_0}
\label{psime1}
\end{equation}
\begin{equation}
\ket{\psi_\nu^{\mathrm{N}+1}} 
= \hat{O}^\dagger_\nu \ket{\psi^{\mathrm{N}}_0}
= \sum\limits _{i}C^{*}_{i\nu}\hat{a}^\dagger_{i}\ket{\psi^{\mathrm{N}}_0}
\label{psima1}
\end{equation}
Here, $\hat{a}_{i}$ ($\hat{a}^\dagger_{i}$) is the annihilation (creation) operator for an electron in the spin-orbital $\ket {\phi_i} = \ket{\varphi_{p}} \otimes \ket{\sigma}$  $\left(\sigma = \alpha, \beta \right)$, and $\left\{C_{i\nu}\right\}$ are the set of coefficients to be determined. Optimizing the energy of the state $\psi_{\nu}^{\mathrm{N}\mp1}$ with respect to the parameters $\left\{C_{i\nu}\right\}$ and subtracting the energy of $\psi_0^\mathrm{N}$, gives the EKT equations as a generalized eigenvalue problem, namely,
\begin{equation}
\mathrm{T^{\mp}C} = \mathrm{\Gamma^{\mp} C \omega}
\label{ekt}
\end{equation}

In this context, the metric matrix $\Gamma^{-}$ is the 1RDM ($\Gamma$), consisting of the ONs along the diagonal and zeros in the off-diagonal elements in the NO representation. Conversely, $\Gamma^{+}=1-\Gamma$, with 1 representing the unit matrix. On the other hand, $\omega$ corresponds to the IPs when the transition matrix elements are given by \cite{Day1974, Smith1975}
\begin{equation}
T^{-}_{ij} = \bra{\psi^\mathrm{N}_0} \hat{a}_{i}^{\dagger} \comm{\hat{\mathcal{H}}_{el}}{\hat{a}_{j}} \ket{\psi^\mathrm{N}_0} 
\end{equation}
or the EAs if the transition matrix elements are given by
\begin{equation}
T^{+}_{ji} = \bra{\psi^\mathrm{N}_0} \hat{a}_{i} \comm{\hat{\mathcal{H}}_{el}}{\hat{a}_{j}^{\dagger}} \ket{\psi^\mathrm{N}_0} 
\end{equation}

It is not challenging to show that the transition matrix elements $T^{\mp}_{ij}$ are related to $\lambda_{ij}$. For instance, 
\begin{equation}
T^{-}_{ij} = -(n_{j}H_{ij} + 2\sum\limits_{klm} D_{iklm} \left\langle lm|jk \right\rangle) = - \lambda_{ij}
\end{equation}
By employing a spin-restricted theory, Eq. (\ref{ekt}) can be modified through canonical orthonormalization using $\Gamma^{-1/2}$. Thus, the diagonalization of the matrix $\mathbf{\omega}$, whose elements are:
\begin{equation}
\omega_{qp}=-\frac{\lambda_{qp}}{\sqrt{n_{q}n_{p}}}\label{niu}
\end{equation}
yields IPs as eigenvalues. A similar transformation is also valid for EAs.

The IPs and EAs obtained from the EKT with NOFs have undergone extensive comparisons with experimental data and other theoretical calculations \cite{Pernal2005a, Leiva2006, Leiva2007a, Piris2008b, Piris2012, Matxain2012, Zarkadoula2012, Piris2013, Matxain2013a, Piris2015, DiSabatino2015, Piris2016}. Recent improvements, such as incorporating electron screening into EKT, have successfully addressed band gap opening in both weakly and strongly correlated systems \cite{DiSabatino2022, DiSabatino2023}, albeit with some remaining inaccuracies in full photo-emission spectra.

The overall agreement of EKT combined with NOF approximations is generally satisfactory, particularly for lower IPs. However, this agreement tends to diminish for higher IPs compared to the accuracy observed for the lowest IPs. Conversely, the calculation of EAs is notably more challenging through EKT. Presently, EKT provides an overall unsatisfactory description of EAs. Nonetheless, to evaluate the estimation of EAs, one can utilize the inverse of the IP of the corresponding anionic species, calculated at the experimental geometry of the neutral species. The calculated EKT EAs using this strategy exhibit good agreement with experimental values, presenting an improvement over the KT approach \cite{Piris2012}.

\subsection{Neutral excitations: Extended random phase approximation (ERPA)}

In the 1RDM framework, time-dependent 1RDMFT in its adiabatic linear response formulation has been developed \cite{Pernal2007a, Pernal2007, Pernal2007b, Giesbertz2008, Giesbertz2009, Giesbertz2010a, VanMeer2014, VanMeer2017} to calculate the energies of excited states, however, a solid foundation for a dynamic 1RDMFT is still an open challenge \cite{Pernal2016}. On the other hand, an ensemble version of 1RDMFT has recently been proposed \cite{Schilling2021} to calculate the energies of selected low-lying excited states, although it will require more efficient numerical minimization schemes for its future success \cite{Liebert2022}.

In a recent work \cite{Lew-Yee2023d}, we extended PNOFs to excited states by integrating their reconstructed 2RDMs with the extended random phase approximation (ERPA) \cite{Gambacurta2008, Gambacurta2009}. Specifically, we adopted the formulation by Chatterjee and Pernal \cite{Chatterjee2012}, relying on the 1RDM and 2RDM of the ground state. This method is elegantly derived from the formally exact Rowe's excitation operator equation-of-motion \cite{Rowe1968}. In the subsequent discussion, we refrain from using the superscript N, given that the number of particles remains constant.

In this scenario, we define the expectation value of a double commutator involving the excitation operator for a system governed by a Hamiltonian $\hat{\mathcal{H}}_{el}$ as follows:
\begin{equation}
\bra{\psi_0} \comm{\delta \hat{O}_\nu}{\comm{\hat{\mathcal{H}}_{el}}{\hat{O}_\nu^\dagger}} \ket{\psi_0}=\omega \bra{\psi_0} \comm{\delta \hat{O}_\nu}{\hat{O}_\nu^\dagger} \ket{\psi_0}
\label{eq:eom}
\end{equation}
Here, $\omega$ corresponds to the excitation energy $E_{\nu} - E_0$, and $\hat{O}_\nu^\dagger$ is the excitation operator that, when applied to the ground state $\ket{\psi_0}$, produces the excited state $\ket{\psi_\nu}$:
\begin{equation}
    \hat{O}^\dagger_\nu \ket{\psi_0} = \ket{\psi_\nu} \>\>
\end{equation}
Furthermore, $\hat{O}_\nu$ deexcitates from $\ket{\psi_\nu}$ to $\ket{\psi_0}$ and satisfies the consistency condition to ensure the orthogonality of the ground and excited states:
\begin{equation}
    \hat{O}_\nu \ket{\psi_0} = 0 \>\>
    \label{eq:consistent-condition}
\end{equation}

Different approximations arise from the use of various excitation operators. In its most basic form, incorporating solely single non-diagonal excitations, we obtain the ERPA0 approximation. By additionally incorporating the single diagonal excitations, we arrive at the ERPA1 approximation. Regrettably, both approaches violate consistency condition (\ref{eq:consistent-condition}), which leads to a degradation in the excitation energies. Finally, with the inclusion of double diagonal excitations instead of single diagonal excitations, we reach ERPA2, which possesses the significant advantage of enforcing (\ref{eq:consistent-condition}), at least for two-electron systems. In correspondence with the preceding, we have the following approximations for the excitation operator $\hat{O}_\nu^\dagger$:
\begin{equation}
\hat{O}^\dagger_\nu [\mathrm{0}]
= \sum_{p>q} X_{pq} \left( a_{p_{\alpha}}^\dagger a_{q_{\alpha}} 
+ a_{p_{\beta}}^\dagger a_{q_{\beta}} \right) 
+ \sum_{p>q} Y_{pq} \left( a_{q_{\alpha}}^\dagger a_{p_{\alpha}}
+ a_{q_{\beta}}^\dagger a_{p_{\beta}} \right)
\label{eq:exop0}
\end{equation}
\begin{equation}
\hat{O}^\dagger_\nu [\mathrm{1}] = \hat{O}^\dagger_\nu [\mathrm{0}]
+ \sum_{p} Z_{p} \left( a_{p_{\alpha}}^\dagger a_{p_{\alpha}} + a_{p_{\beta}}^\dagger a_{p_{\beta}} \right)
\label{eq:exop1}
\end{equation}
\begin{equation}
\hat{O}^\dagger_\nu [2] = \hat{O}^\dagger_\nu [0]
+ \sum_{pq} W_{pq} \left( a_{p_{\beta}}^\dagger a_{q_{\beta}} a_{p_{\alpha}}^\dagger a_{q_{\alpha}} \right) 
\label{eq:exop2}
\end{equation}
Here, $X_{pq}$, $Y_{pq}$, $Z_{p}$, and $W_{pq}$ represent coefficients to be determined. By calculating the variation of the adjoint of the corresponding excitation operator and substituting it into Eq. (\ref{eq:eom}), the resultant systems of equations to be solved are acquired. These can be regarded as generalized eigenvalue problems, where the coefficients are determined by the 1RDM and 2RDM. For instance, the following generalized eigenvalue problem is formulated in matrix form for ERPA0:
\begin{multline}
\qquad\quad
\begin{pmatrix} A_{rspq} & B_{rspq} \\ B_{rspq} & A_{rspq} \end{pmatrix}
\begin{pmatrix} X_{pq} \\ Y_{pq} \end{pmatrix}
= \omega 
\begin{pmatrix} \Delta N_{rspq} & 0 \\ 0 & -\Delta N_{rspq} \end{pmatrix}
\begin{pmatrix} X_{pq} \\ Y_{pq} \end{pmatrix}
\label{eq:ERPA0}
\end{multline}
where
\begin{multline}
    A_{rspq} = h_{sq} \delta_{pr} (n_p - n_s) + h_{pr} \delta_{sq} (n_q - n_r)
    + \sum_{tu} (\braket{qt}{su} - \braket{qt}{us}) \mathrm{D}_{ptru}^{\alpha\alpha\alpha\alpha}\\
    + \sum_{tu} \braket{qt}{su} \mathrm{D}_{ptru}^{\alpha\beta\alpha\beta} + \sum_{tu} \braket{qt}{us} \mathrm{D}_{ptur}^{\alpha\beta\alpha\beta}
    + \sum_{tu} (\braket{pt}{ru} - \braket{pt}{ur}) \mathrm{D}_{qtsu}^{\alpha\alpha\alpha\alpha}\\
    + \sum_{tu} \braket{pt}{ru} \mathrm{D}_{qtsu}^{\alpha\beta\alpha\beta} + \sum_{tu} \braket{pt}{ur} \mathrm{D}_{qtus}^{\alpha\beta\alpha\beta}
    + \sum_{tu} \braket{ps}{ut}\mathrm{D}_{qrtu}^{\alpha\alpha\alpha\alpha} - \sum_{tu} \braket{ps}{tu} \\ 
    \mathrm{D}_{qrtu}^{\alpha\beta\alpha\beta}
    + \sum_{tu} \braket{qr}{ut}\mathrm{D}_{pstu}^{\alpha\alpha\alpha\alpha} 
    - \sum_{tu} \braket{qr}{tu}\mathrm{D}_{pstu}^{\alpha\beta\alpha\beta}
    + \delta_{qs} \sum_{tuv} \braket{pt}{vu}\mathrm{D}_{rtuv}^{\alpha\alpha\alpha\alpha}\\ 
    - \delta_{qs} \sum_{tuv} \braket{pt}{uv}\mathrm{D}_{rtuv}^{\alpha\beta\alpha\beta}
    + \delta_{pr} \sum_{tuv} \braket{qt}{vu}\mathrm{D}_{stuv}^{\alpha\alpha\alpha\alpha} 
    - \delta_{pr} \sum_{tuv} \braket{qt}{uv}\mathrm{D}_{stuv}^{\alpha\beta\alpha\beta}\>,
    \nonumber
    \label{eq:A}
\end{multline}
\begin{equation}
    B_{rspq} = A_{rsqp} \>,\quad   
    \Delta N_{rspq} = (n_s - n_r) \delta_{rp}\delta_{sq} \qquad\qquad\qquad\qquad\qquad\qquad\quad
    \nonumber
\end{equation}
Similar equations are obtained for the ERPA1 and ERPA2 approximations. The details can be found in Ref. \cite{Lew-Yee2023d}.

The results obtained so far demonstrate that coupling electron-pairing-based NOFs, Sec. \ref{sec:pairNOF}, with the ERPAs is a promising approach. Indeed, a good agreement was achieved with excited state calculations using the Full Configuration Interaction (FCI) method. As expected, transitioning from ERPA0 to ERPA1 and ERPA2 improved the results. ERPA0 exhibited inaccuracies concerning avoided crossings in the studied systems, and although ERPA1 enhanced the outcomes, ERPA2 was required for an accurate description. Regarding the tested functionals, PNOF5 appears to be sufficient for small molecules, but as the system size increases, electronic pair correlation becomes crucial, and PNOF7 and GNOF yielded superior results. It is noteworthy that the scaling of excited state calculations with respect to the system size is sixth order, comparable to standard TD-DFT, but with substantially improved results.

\section{Closing Remarks}

Löwdin's introduction of NOs and their application in describing ground states of two-electron systems laid the foundation for NOFT. Currently, NOFT is actively evolving, providing a middle ground in cost between multireference methods and conventional density functionals. Notably, NOFT exhibits heightened accuracy in comparison to electron density-dependent alternatives, especially in systems characterized by substantial non-dynamic correlation.

This chapter has outlined the theoretical framework for approximate NOFs, placing emphasis on the challenges and advancements in their formulation, particularly in addressing N-representability functional issues. Various two-index reconstructions for the 2RDM have been reviewed, and a detailed analysis has delved into NOFs grounded in electron pairing, with specific attention given to PNOF5, PNOF7, and the Global NOF, offering a more versatile approach to addressing both static and dynamic electron correlation components.

The extension of NOFs to multiplets while conserving total spin, supported by open-source implementations like \href{http://github.com/DoNOF}{DoNOF}, is highlighted. The chapter provides a comprehensive examination of optimization procedures for single-point calculations. While optimization with respect to ONs has advanced significantly and can now be conducted efficiently, optimizing with respect to NOs has improved but still necessitates substantial enhancements for NOF-based methods to be competitive with density functional approximations. The iterative diagonalization method proves computationally efficient for larger systems, and a hybrid approach combining it with orbital rotations streamlines NO optimization.

The current most efficient optimization process still utilizes the embedded loop algorithm, optimizing the ONs while keeping the NOs fixed, and vice versa. Enhanced by strategies like parallelization, integral-direct formalism, and the resolution of identity approximation, existing codes, including DoNOF, have significantly improved efficiency. This progress facilitates the application of NOF approximations to larger molecular systems, showcasing the codes' capability to tackle complex and chemically relevant challenges.

The chapter encompasses sections on geometry optimization and AIMD. Analytical gradients concerning nuclear motion can be readily acquired through a straightforward evaluation, obviating the necessity for linear response theory. Calculations substantiate that equilibrium geometries and harmonic vibrational frequencies exhibit satisfactory agreement with accurate theoretical methods and available experimental data. GNOF-AIMD marks a significant advancement, positioning it as a recommended technique for exploring the real-time evolution of complex electronic problems, particularly valuable for observing the dynamic changes in NOs during alterations in bond patterns.

The chapter concludes by presenting the extension of NOFs to both charged and excited states, exhibiting their versatility in capturing diverse electronic phenomena. The EKT effectively describes ionization potentials, while challenges persist in calculating electron affinities. Coupling electron-pairing-based NOFs with ERPA proves promising for studying neutral excitations. Assessments indicate that PNOF5 suffices for small molecules, while GNOF yields superior results as the system size increases, with comparable scaling in excited state calculations to standard TD-DFT but with substantially improved outcomes.

\subsection*{Acknowledgments}

Financial support comes from MCIU (PID2021-126714NB-I00) and Eusko Jaurlaritza (IT1584-22). The author thanks for technical and human support provided by SGIker (UPV/EHU/ERDF,EU), for the allocation of computational resources provided by the Scientific Computing Service.

\newpage


\end{document}